\documentclass[prl,aps,twocolumn,superscriptaddress]{revtex4-2}
\usepackage[utf8]{inputenc}
\usepackage{amsmath, amssymb, amsbsy}
\usepackage{color}
\usepackage{array}
\usepackage{graphicx}
\usepackage[percent]{overpic}
\usepackage{extarrows}
\usepackage{appendix}
\usepackage{xcolor}
\usepackage{braket}
\usepackage[export]{adjustbox}
\usepackage{multirow}

\usepackage[colorlinks=true,linkcolor=blue,citecolor=blue,urlcolor=blue,bookmarks=false]{hyperref}

\pagecolor{white}

\begin{document}

\title{Phase diagram of the XXZ pyrochlore model from pseudo-Majorana functional renormalization group}

\date{\today}

\author{Yannik Schaden}
\affiliation{Dahlem Center for Complex Quantum Systems and Institut f\"ur Theoretische Physik, Freie Universit\"at Berlin, Arnimallee 14, 14195 Berlin, Germany}
\affiliation{Helmholtz-Zentrum Berlin f\"ur Materialien und Energie, Hahn-Meitner-Platz 1, 14109 Berlin, Germany}

\author{Mat\'ias G. Gonzalez}
\affiliation{Dahlem Center for Complex Quantum Systems and Institut f\"ur Theoretische Physik, Freie Universit\"at Berlin, Arnimallee 14, 14195 Berlin, Germany}
\affiliation{Helmholtz-Zentrum Berlin f\"ur Materialien und Energie, Hahn-Meitner-Platz 1, 14109 Berlin, Germany}

\author{Johannes Reuther}
\affiliation{Dahlem Center for Complex Quantum Systems and Institut f\"ur Theoretische Physik, Freie Universit\"at Berlin, Arnimallee 14, 14195 Berlin, Germany}
\affiliation{Helmholtz-Zentrum Berlin f\"ur Materialien und Energie, Hahn-Meitner-Platz 1, 14109 Berlin, Germany}

\begin{abstract}
We calculate the magnetic phase diagram of the spin-$1/2$ nearest neighbor XXZ pyrochlore model using the pseudo-Majorana functional renormalization group in the temperature flow formalism. Our phase diagram as a function of temperature and coupling ratio, allowing both longitudinal and transverse couplings to be ferromagnetic and antiferromagnetic, reveals a large non-magnetic regime at low temperatures, which includes the quantum spin ice phase near the antiferromagnetic Ising model, as well as the antiferromagnetic Heisenberg and XY models. We are able to detect magnetic phase transitions via critical finite size scaling down to temperatures two orders of magnitude smaller than the spin interactions, demonstrating the remarkably good performance of our method upon approaching the ground state. Specifically, the low temperature transition from the zero-flux quantum spin ice phase into the transverse ferromagnetic phase shows very good agreement with previous quantum Monte Carlo results. Comparing our findings with classical results, we identify a quantum order-by-disorder effect near the antiferromagnetic XY model. In magnetically disordered regimes, we find characteristic patterns of broadened pinch points in the spin structure factor and investigate their evolution when approaching magnetically ordered phases. We also compute linear responses to lattice symmetry breaking perturbations and identify a possible lattice nematic ground state of the antiferromagnetic XY model.
\end{abstract}

\maketitle

\section{Introduction} \label{sec:introduction}
The pyrochlore lattice is one of the most studied three-dimensional networks in quantum magnetism. The ongoing research interest in pyrochlore systems is motivated by the wealth of material realizations~\cite{RevModPhys.82.53,annurev:10.1146,GREEDAN2006444} and the intriguing magnetic properties that this lattice can give rise to~\cite{Gingras_2014,PhysRevX.1.021002,Savary_2017,balents2010}. Already the {\it classical} nearest neighbor antiferromagnetic Ising model on the pyrochlore lattice, host of the famous spin ice phase, produces non-trivial magnetic effects such as residual entropy and fractional excitations~\cite{castelnovo2008,ramirez1999}. Even more strikingly, when adding small transverse $xx$ and $yy$ couplings to the Ising model, it can be shown perturbatively that the system realizes a quantum spin liquid captured by an emergent quantum electrodynamics theory~\cite{huse2003,PhysRevB.86.075154,PhysRevB.69.064404,PhysRevLett.108.037202}. Identifying the signatures of this {\it quantum} spin ice phase in magnetic materials, such as dipolar-octopolar compounds~\cite{PhysRevX.12.021015,bhardwaj2022,PhysRevX.14.011005,sibille_quantum_2020,gao_experimental_2019,gao2024}, is a particularly active area of research. Away from the perturbative regime of quantum spin ice, the spin-1/2 nearest neighbor pyrochlore Heisenberg model has also attracted significant attention as a highly frustrated pyrochlore spin model. Despite very rare material realizations~\cite{clark2014,PhysRevMaterials.1.071201}, the spin-1/2 pyrochlore Heisenberg model has a long history in terms of theoretical investigations aiming to identify novel strongly fluctuating quantum phases~\cite{PhysRevLett.80.2933,PhysRevLett.121.067201,PhysRevX.7.041057,PhysRevLett.116.177203,PhysRevLett.126.117204,PhysRevB.65.024415,doi:10.1143/JPSJ.67.4022,PhysRevX.9.011005,10.21468/SciPostPhys.12.5.156,harris_ordering_1991,PhysRevLett.131.096702}.

In this paper, we numerically investigate the nearest neighbor spin-1/2 XXZ pyrochlore model which interpolates between different interesting limits including the aforementioned Ising and Heisenberg systems. Specifically, we study the full phase diagram where the longitudinal $zz$ and transverse $xx$/$yy$ couplings can both be ferromagnetic and antiferromagnetic. While the classical version of this model is well understood~\cite{PhysRevX.7.041057}, the quantum system is much less explored mostly due to a lack of numerical methods that are capable of treating three dimensional spin networks in combination with highly frustrated interactions. Some progress has been achieved with quantum Monte Carlo techniques in the case of unfrustrated transverse interactions, which could, e.g., resolve the transition of the quantum spin ice phase into an ordered magnet upon increasing these couplings~\cite{PhysRevLett.115.077202,PhysRevLett.100.047208,PhysRevResearch.2.042022}, revealing a quantum order-by-disorder effect. However, in other parts of the phase diagram such as around the XY model the effects of quantum fluctuations beyond mean-field have been rarely studied~\cite{PhysRevLett.121.067201,PhysRevB.109.184421}.

Here, we apply the pseudo-fermion functional renormalization group (PMFRG) method~\cite{PhysRevB.103.104431,10.21468/SciPostPhys.12.5.156,Muller_2024} in its temperature-flow ($T$-flow) variant~\cite{PhysRevB.109.195109} to the XXZ pyrochlore model. In a number of recent applications, the PMFRG has proven to be a flexible and accurate technique to study quantum spin systems even in the case of complex and highly frustrated spin interactions~\cite{10.21468/SciPostPhys.12.5.156,PhysRevB.109.195109,PhysRevB.103.104431,PhysRevLett.130.196601,PhysRevB.104.L220408,niggemann_decorated,hagymasi2024,bippus2024,PhysRevB.109.144411}. In the version applied here, two method extensions of the PMFRG as it has been originally introduced in Refs.~\cite{PhysRevB.103.104431,10.21468/SciPostPhys.12.5.156} are combined, the $T$-flow scheme of Ref.~\cite{PhysRevB.109.195109} and the generalization from isotropic Heisenberg to XXZ interactions~\cite{PhysRevB.109.195109,bippus2024}. Particularly, we take advantage of the increased numerical efficiency of the $T$-flow scheme as compared to the conventional $\Lambda$-flow scheme which allows us to calculate the system's phase diagrams as a function of temperature with moderate numerical efforts and which is capable of detecting magnetic phase transitions even at small temperatures~\cite{PhysRevB.109.195109}. Given the novelty of our method, in addition to the actual physics questions that we will address below, this work also aims to assess the accuracy and performance of the technique, by comparing our findings with literature results.

Our investigations can be regarded as a continuation of Ref.~\cite{PhysRevB.109.184421} where the XYZ model on the pyrochlore lattice is studied with the pseudo-fermion functional renormalization group (PFFRG) approach. The PFFRG is a precursor of the PMFRG that has been mostly applied to spin models at zero temperature (including investigations of pyrochlore magnets~\cite{PhysRevB.107.214414,gomez2024,PhysRevB.105.054426,PhysRevB.106.235137,PhysRevB.109.184421,PhysRevX.9.011005,ghosh2019,PhysRevMaterials.1.071201}). While PFFRG and PMFRG differ in the way the spin operators are expressed in terms of fermions (complex fermions versus Majorana fermions) and therefore appear formally similar, they are still independent approaches with different approximations of magnetic correlation functions~\cite{Muller_2024}. The advantage of the PMFRG over the PFFRG is the absence of possible spurious effects from unphysical fermionic states~\cite{PhysRevB.106.235113} in the employed Majorana spin representation, the accessibility of finite temperatures~\cite{PhysRevB.103.104431} and the capability of detecting magnetic phase transitions via scaling collapse~\cite{10.21468/SciPostPhys.12.5.156}. On the other hand, our PMFRG investigations of an XXZ model make use of global spin rotation symmetry around the $z$-axis. Thus, a further generalization of the PMFRG for general anisotropic XYZ interactions without continuous spin symmetries requires further method development which goes beyond the scope of this work.

\begin{figure*}
	\begin{center}
		\includegraphics[width=1.00\textwidth]{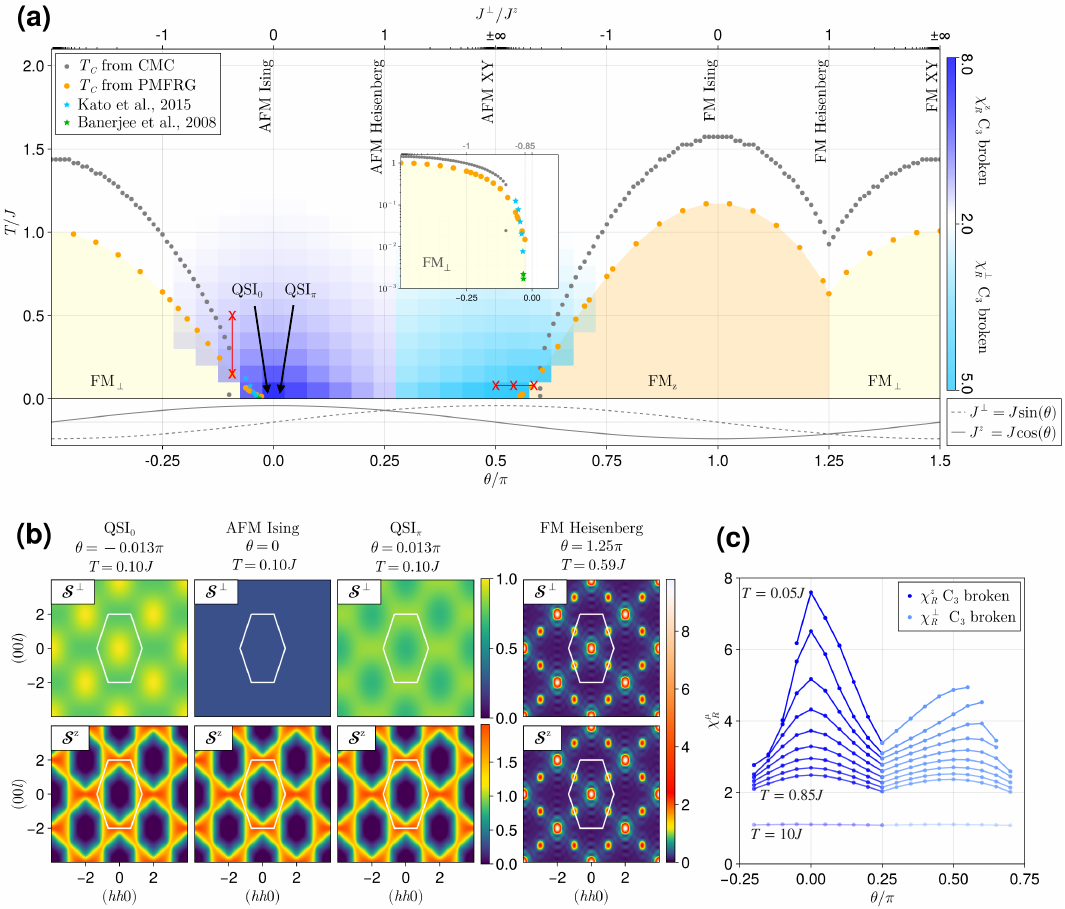}
	\caption[Phase Diagram]{(a) $\theta$-$T$ phase diagram of the pyrochlore XXZ model with magnetic phase transitions from PMFRG (orange points) and from classical Monte Carlo with three-components spins normalized as $|\boldsymbol{S}_i|=1$ (gray points). The inset shows an enlarged version of the region near the transition between FM$_\perp$ and QSI$_0$ phases with a logarithmic temperature axis. The blue and green stars indicate the phase transitions from Ref.~\cite{PhysRevLett.115.077202} and~\cite{PhysRevLett.100.047208}, respectively. Dark blue and light blue squares correspond to the color scale on the right indicating the linear responses $\chi_R^z$ and $\chi_R^\perp$ of the system to a perturbation that breaks C$_3$ lattice rotation symmetry [see Sec.~\ref{sec:nematic}]. Red crosses mark points, at which structure factors are shown in Fig.~\ref{fig:add_SF}. (b) Spin structure factors in the $(hhl)$ plane [i.e. at wave vectors $\boldsymbol{Q}=2\pi(h,h,l)$] at selected angles $\theta$. The longitudinal spin structure factor at the Ising point shows a constant signal $\mathcal{S}^{z}(\boldsymbol{Q})=1/4$, which is taken from analytical considerations~\cite{PhysRevB.106.235113}. White hexagons indicate the extended Brillouin zone. (c) Linear responses $\chi_R^z$ and $\chi_R^\perp$ for C$_3$ lattice rotation symmetry breaking as a function of $\theta$ from $T = 0.05J$ to $T = 0.85J$ (in steps of $0.1J$). The data is the same as for the light and dark blue squares in (a). In the high temperature limit the system does not respond to small perturbations such that $\chi_R^{\perp/z}\approx1$, as illustrated for $T=10J$.}
	\label{fig:phasediagram}
	\end{center}
\end{figure*}

The main result of our work is the full magnetic phase diagram of the spin-1/2 XXZ pyrochlore model as a function of temperature, see Fig.~\ref{fig:phasediagram}(a). We determine the extent of the non-magnetic phase around the antiferromagnetic Ising limit and find excellent agreement with quantum Monte Carlo when the transverse couplings are ferromagnetic (unfrustrated). Interestingly, we find that the $T$-flow scheme correctly detects magnetic phase transitions via a finite-size scaling collapse even at temperatures two orders of magnitudes below the energy scale of the bare couplings, i.e., far beyond the high-temperature regime where the approach is perturbatively controlled~\cite{PhysRevB.103.104431}. The large non-magnetic low-temperature phase which we identify also includes the antiferromagnetic XY model and even remains stable when adding small ferromagnetic longitudinal interactions in the latter model. Nevertheless, the obtained non-magnetic low-temperature regime is smaller than in the corresponding classical model, indicating quantum order-by-disorder effects. To characterize the magnetic correlations, we discuss momentum resolved spin structure factors in various parts of the phase diagram. The non-magnetic regime is further studied by calculating the system's response to lattice symmetry breaking perturbations. This investigation is motivated by a number of recent works which found that the ground state of the antiferromagnetic pyrochlore Heisenberg model exhibits broken lattice symmetries~\cite{PhysRevLett.126.117204,PhysRevX.11.041021,PhysRevB.105.054426,PhysRevLett.131.096702}. Our $T$-flow PMFRG results also indicate an enhanced response of the antiferromagnetic pyrochlore Heisenberg model to lattice rotation symmetry breaking perturbations. Interestingly, this response increases even further upon approaching the antiferromagnetic XY limit. This suggests a lattice nematic phase as a promising ground state candidate of the antiferromagnetic XY model.

The rest of the paper is structured as follows. In Sec.~\ref{sec:model} we introduce the XXZ pyrochlore model and summarize known results of its phase diagram. Sec.~\ref{sec:methods} provides a brief introduction to PMFRG, deferring technical details to the Appendix and to other literature. In Sec.~\ref{sec:results} we present and discuss our findings, put them into the context of existing results and compare them to predictions from the classical XXZ pyrochlore model. The paper ends with a conclusion in Sec.~\ref{sec:summary}.

\section{Model}\label{sec:model}
In this work we investigate the XXZ model on the pyrochlore lattice whose Hamiltonian reads as
\begin{align}
    H_\mathrm{XXZ} = \sum_{\langle i,j \rangle} \big[ J^z S_i^z S_j^z + J^\perp \left( S_i^x S_j^x + S_i^y S_j^y \right) \big],\label{eq:XXZhamiltonian}
\end{align}
where $S_i^\mu$ is the $\mu$-th component of the spin-1/2 operator on site $i$ and $\langle i,j \rangle$ are nearest neighbor pairs of sites.  Throughout this work we parametrize the couplings as $J^\perp = J \, \mathrm{sin}(\theta)$ and $J^z = J \, \mathrm{cos}(\theta)$. In the following, we briefly summarize the predictions and results for this model in well investigated regions of the phase diagram, as reported in the literature.

At $\theta=0$ (antiferromagntic $J^z>0$ and $J^\perp=0$) the system realizes classical spin ice, a classical spin liquid with an extensive ground state degeneracy and dipolar spin correlations~\cite{PhysRevB.109.174421,PhysRevLett.81.4496,doi:10.1126/science.1064761,PhysRevLett.93.167204,yan2023} that is most notably realized in the materials $\text{Ho}_2\text{Ti}_2\text{O}_7$~\cite{PhysRevLett.87.047205,PhysRevLett.79.2554} and $\text{Dy}_2\text{Ti}_2\text{O}_7$~\cite{Lago_2007,ramirez1999}. As a characteristic spectroscopic signature of the dipolar correlations, classical spin ice exhibits sharp pinch points in the spin structure factor~\cite{PhysRevLett.93.167204}. For small $|\theta|\ll1$ ($|J^\perp|\ll J^z$) where transverse interactions can be treated perturbatively the system enters a quantum spin ice phase, a U(1) quantum spin liquid described by an emergent electrodynamics theory~\cite{huse2003,PhysRevB.86.075154,PhysRevB.69.064404}. For $J^\perp\lesssim0$ the system realizes 0-flux quantum spin ice (QSI$_0$) characterized by a ground state with zero emergent magnetic flux while $J^\perp\gtrsim0$ gives rise to $\pi$-flux quantum spin ice (QSI$_\pi$) where an emergent magnetic flux of $\pi$ pierces through each elementary hexagon of the pyrochlore lattice~\cite{PhysRevLett.132.066502,sanders2024}. Upon increasing $|J^\perp|$ in the unfrustrated QSI$_0$ regime, the quantum spin liquid phase quickly becomes unstable, giving way to ferromagnetic long-range order in the $x$-$y$ plane, spontaneously breaking the U(1) spin rotation symmetry around the $z$-axis. The zero-temperature transition between both phases has been located at $\theta=-0.033\pi$ ($J^\perp/J^z = -0.1$) by quantum Monte Carlo~\cite{PhysRevLett.115.077202,PhysRevLett.100.047208} and at $\theta=-0.03\pi$ $(J^\perp/J^z = -0.096)$ by gauge mean field theory~\cite{PhysRevLett.132.066502}. Both values are closer to the pure Ising limit than $\theta=-0.1\pi$ ($J^\perp/J^z=-0.33$) which is the transition of the corresponding classical model~\cite{PhysRevX.7.041057,PhysRevLett.116.217201,PhysRevB.83.094411,chung2024,lozano2024}, indicating a quantum order-by-disorder mechanism~\cite{gomez2024}. Increasing $J^\perp$ in the frustrated QSI$_\pi$ regime ($J^\perp>0$) poses severe challenges to most numerical methods. Nevertheless, there is growing consensus that the quantum model remains non-magnetic at least up to the Heisenberg point at $\theta=\pi/4$ ($J^\perp=J^z>0$)~\cite{PhysRevB.109.184421} with additional claims that the ground state remains QSI$_\pi$ in the entire regime up to $\theta=\pi/4$~\cite{PhysRevLett.121.067201,benton2020}.

The antiferromagnetic spin-1/2 Heisenberg model ($\theta=\pi/4$) has been a separate subject of investigations over the last decades with a plethora of different outcomes concerning its ground state properties~\cite{PhysRevLett.80.2933,PhysRevLett.121.067201,PhysRevX.7.041057,PhysRevLett.116.177203,PhysRevLett.126.117204,PhysRevB.65.024415,doi:10.1143/JPSJ.67.4022,PhysRevX.9.011005,10.21468/SciPostPhys.12.5.156,harris_ordering_1991,PhysRevLett.131.096702}. While early on, the system has been predicted to be a potential candidate for quantum spin liquid behavior~\cite{PhysRevLett.80.2933}, more recent works identify lattice symmetry breaking~\cite{PhysRevLett.126.117204,PhysRevX.11.041021,PhysRevB.105.054426,PhysRevLett.131.096702}. While these tendencies may be interpreted as an indication for a valence-bond crystal, suggested to consist of hard-hexagon tilings~\cite{PhysRevLett.131.096702}, other works found that lattice symmetry breaking gives rise to dimensional reduction and an effective two-dimensional quantum spin liquid~\cite{pohle2023groundstates12pyrochlore}.

Upon further increasing $\theta$, the antiferromagnetic XY model is reached at $\theta=\pi/2$ ($J^z=0$, $J^\perp>0$). The classical version of this model is well-known to exhibit a thermal order-by-disorder mechanism which selects collinear spin-ice states out of the continuously degenerate ground state manifold~\cite{PhysRevB.58.12049,PhysRevLett.80.2929,PhysRevX.7.041057}. On the other hand, the quantum XY model is much less explored with predictions ranging from a spin nematic state~\cite{PhysRevLett.121.067201} to a QSI$_\pi$ state~\cite{benton2020}. Upon increasing $\theta$ away from the XY model, the longitudinal couplings $J^z<0$ drive the system into a ferromagnetic Ising phase. By a combination of different variational and exact diagonalization methods the transition between both phases has been found at $\theta=0.613\pi$ ($J^\perp/J^z=-2.7$) in Ref.~\cite{PhysRevLett.121.067201}. This value is larger than the transition of the corresponding classical model at $\theta=0.602\pi$ ($J^\perp/J^z=-3$)~\cite{PhysRevLett.121.067201}, implying that quantum fluctuation partially melt the ferromagnetic Ising order. This is in contrast to the other end of the non-magnetic phase discussed before, where the magnetically ordered phase grows in the quantum case, showing a quantum order-by-disorder effect.

\section{Numerical Methods} \label{sec:methods}
In the following, we provide a brief overview of the numerical methods employed. The full quantum mechanical phase diagram, the spin structure factors and the responses to symmetry breaking perturbations are computed using PMFRG. Additionally, we apply classical Monte Carlo simulations to identify effects with a quantum origin.

\subsection{PMFRG} \label{sec:pmfrg}
The central methodological step in PMFRG is the reformulation of spin operators $S_i^\mu$ in terms of Majorana operators $\eta_i^\mu$ via
\begin{align}
    S_i^x = -\mathrm{i}\eta_i^y\eta_i^z, \qquad
    S_i^y = -\mathrm{i}\eta_i^z\eta_i^x, \qquad
    S_i^z = -\mathrm{i}\eta_i^x\eta_i^y, \label{eq:spinmajorana}
\end{align}
where $\eta_i^\mu$ satisfy the Clifford algebra relations $\{\eta_i^\mu,\eta_j^\nu\} = \delta_{ij}\delta_{\mu\nu}$~\cite{PhysRevLett.69.2142,PhysRevResearch.5.023067,PhysRevB.103.104431,10.21468/SciPostPhys.12.5.156,Muller_2024}. Importantly, this rewriting does not introduce unphysical fermionic states but merely leads to identical copies of the physical Hilbert space. This redundancy, however, does not affect physical observables and, hence, poses no problem for the approach~\cite{PhysRevB.103.104431}. The absence of unphysical states is a key advantage of the PMFRG over the PFFRG~\cite{PhysRevB.81.144410}, its predecessor, where spin operators are expressed in terms of {\it complex} fermions. Furthermore, in contrast to PFFRG, the PMFRG can be straightforwardly applied at finite temperatures. In fact, by virtue of perturbation theory, the PMFRG becomes asymptotically exact at large $T/|J|$~\cite{PhysRevB.103.104431}. 

As is common to all functional renormalization group schemes, a (frequency or momentum) cutoff is introduced in the free fermionic propagator, parametrized by a variable usually called $\Lambda$~\cite{RevModPhys.84.299}. The renormalization group flow starts in the trivial limit where the cutoff completely suppresses the propagator (often defined as the limit $\Lambda\rightarrow\infty$) and all fermionic correlation functions are fully governed by the system's bare interactions. The end of the renormalization group flow is reached when cutoff-free propagators are recovered (typically at $\Lambda=0$). The improvement of the $T$-flow PMFRG approach~\cite{PhysRevB.109.195109} applied here, compared to previous $\Lambda$-flow schemes~\cite{PhysRevB.103.104431,10.21468/SciPostPhys.12.5.156} is that the cutoff parameter $\Lambda$ is implemented as the physical temperature $T$ and not as an artificial parameter. In other words, the renormalization group flow in the $T$-flow approach simulates a physical cooling process. This greatly improves the method's efficiency when computing finite-temperature phase diagrams, as the entire temperature range is covered in a single numerical run.

\begin{figure}
	\begin{center}
		\includegraphics[width=0.48\textwidth]{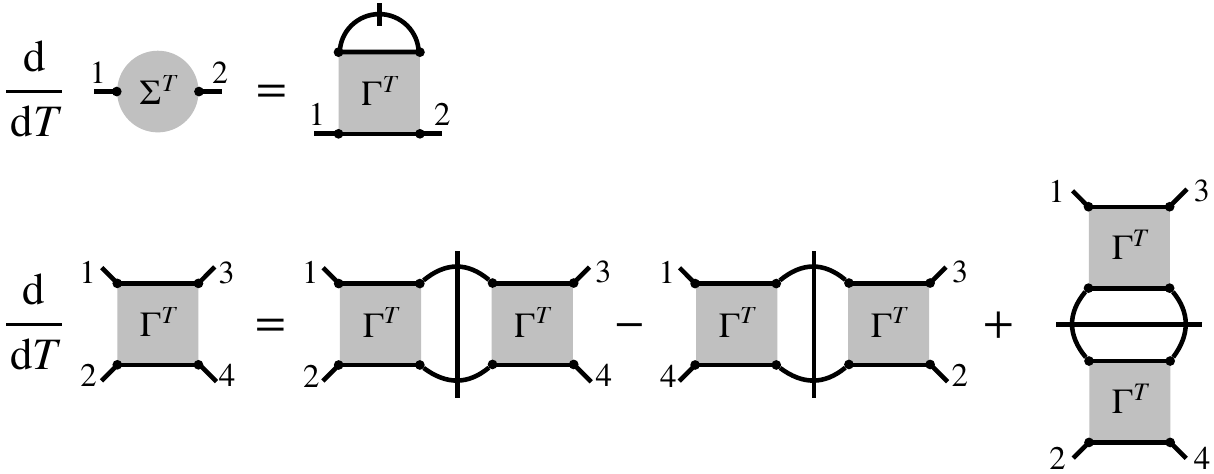}
    \vspace{-16pt}
	\caption[Diagrams PMFRG]{Diagrammatic representation of the PMFRG equations for the self energy $\Sigma^T$ (gray disk) and the two-particle vertex $\Gamma^T$ (gray square) where the label $T$ indicates the temperature dependence. Majorana propagators are illustrated as full lines and a slash indicates a temperature derivative acting on them (a slash cutting two propagators means that the derivative acts on the product of both).}
	\label{fig:diagrams_pmfrg}
	\end{center}
\end{figure}

The renormalization group flow is defined in terms of coupled first order differential equations for the fermionic vertex functions, such as the self-energy $\Sigma$ and the two particle vertex $\Gamma$. These flow equations are formally exact, but they involve arbitrarily high vertex functions, preventing a numerical solution. In order to become solvable, the set of differential equations is truncated, i.e., higher-order vertex functions are set to zero to yield a closed set of equations. Here, we apply the common one-loop truncation where three-particle vertices are neglected (except of certain Katanin contributions~\cite{PhysRevB.81.144410,Muller_2024,katanin2004}). The diagrammatic structure of the resulting differential equations for $\Sigma$ and $\Gamma$ is depicted in Fig.~\ref{fig:diagrams_pmfrg}. We refer the reader to Refs.~\cite{PhysRevB.103.104431,10.21468/SciPostPhys.12.5.156,PhysRevB.109.195109} for an explicit derivation of these equations. Another extension of the PMFRG method employed here concerns the symmetry of the spin interactions. While originally developed for fully spin-isotropic Heisenberg interactions~\cite{PhysRevB.103.104431,10.21468/SciPostPhys.12.5.156}, the renormalization group equations for an XXZ model need to be adapted to account for the reduced symmetry of the spin interactions (see also Refs.~\cite{PhysRevB.109.195109,bippus2024}). In the Appendix~\ref{sec:PMFRG_app} we present details of the derivation of these equations, based on symmetry considerations and present their explicit form in Eqs.~(\ref{eq:floweq1app}) and (\ref{eq:floweq2app}).

The fermionic two-particle vertex allows for a straightforward computation of temperature-dependent spin-spin correlation functions $\langle S_i^\mu S_j^\mu\rangle$ [see Eqs.~(\ref{eq:suscxx}) and (\ref{eq:susczz})] which are the central observables accessible within PMFRG. Crucially, Fourier-transforming the correlators results in the equal-time spin structure factor
\begin{equation}
\mathcal{S}^{\mu}(\boldsymbol{Q})=\langle S^\mu(\boldsymbol{Q})S^\mu(-\boldsymbol{Q})\rangle,
\end{equation}
where $S^\mu(\boldsymbol{Q})$ are the Fourier-transformed spin operators. We note that, since the standard outcome of the PMFRG is the {\it dynamical imaginary frequency} spin structure factor $\mathcal{S}^{\mu}(\boldsymbol{Q},\Omega)$ where $\Omega$ denotes a bosonic Matsubara frequency, the equal time spin structure factor $\mathcal{S}^{\mu}(\boldsymbol{Q})$ is obtained via summation over Matsubara frequencies, $\mathcal{S}^{\mu}(\boldsymbol{Q})=T\sum_\Omega \mathcal{S}^{\mu}(\boldsymbol{Q},\Omega)$.

Below, we use the spin structure factor to illustrate magnetic correlations and to detect magnetic phase transitions. Specifically, denoting the momentum where $\mathcal{S}^{\mu}(\boldsymbol{Q})$ assumes its maximum as $\boldsymbol{Q}^\star$, we may relate the width of this peak to the correlation length $\xi_\mu$ of spin correlations along the $\mu$-direction via~\cite{PhysRevB.109.195109,10.1063/1.3518900}
\begin{align}
    \xi_\mu= \frac{L}{2\pi} \sqrt{\mathcal{S}^{\mu}(\boldsymbol{Q}^\star)/\mathcal{S}^{\mu}(\boldsymbol{Q}^\star+\delta\boldsymbol{q}) - 1}. \label{eq:corrlength}
\end{align}
Here, $\delta \boldsymbol{q}$ denotes the smallest vector in reciprocal space (due to the finite size discretization of $\boldsymbol{Q}$) along the direction of steepest descend away from the peak position. Furthermore, the linear system size $L$ is related to the number of sites $N$ by $N = 4\pi L^3/3$. At a second order phase transition we expect a scaling collapse where $\xi^\mu/L$ is constant as a function of $L$, see Fig.~\ref{fig:corr_lengths}. This determines the transition temperatures of the magnetically ordered regimes in our phase diagram in Fig.~\ref{fig:phasediagram}(a).

In magnetically disordered phases we further probe the system's tendencies to realize lattice-symmetry breaking nematic orders which are characterized by order parameters quadratic in the spin operators $\sim\langle S_i^\mu S_j^\mu\rangle$. To detect such orders we follow Refs.~\cite{PhysRevB.81.144410,PhysRevB.105.054426,PhysRevB.106.235137} and calculate the linear response to small seed fields $\sim S_i^\mu S_j^\mu$ that break the lattice symmetries according to the nematic order to be probed. This amounts to strengthening and weakening interactions on the corresponding bonds [see top of Fig.~\ref{fig:dimer_response}] and evaluating a suited response function. Specifically, denoting sites of strengthened bonds as $k,l$ and those of weakened bonds as $m,n$, the nearest neighbor couplings $J^\mu$ are modified as
\begin{align}
    J^\mu&\rightarrow J_{kl}^\mu \equiv (1+\delta)J^\mu, \notag \\
    J^\mu&\rightarrow J_{mn}^\mu \equiv (1-\delta)J^\mu \label{eq:couplings2}
\end{align}
with $\mu = \perp,z$ and $\delta\ll1$. We then define the temperature dependent linear susceptibility $\chi^\mu_R$ for this perturbation in terms of the response of spin correlations,
\begin{align}
    \chi_R^\mu = \frac{1}{\delta} \left|\frac{\langle S_k^\mu S_l^\mu\rangle-\langle S_m^\mu S_n^\mu\rangle}{\langle S_k^\mu S_l^\mu\rangle+\langle S_m^\mu S_n^\mu\rangle} \right|.\label{eq:response}
\end{align}
The normalizing factor $1/\delta$ in this expression ensures that at large temperatures the nematic susceptibility approaches $\chi_R^\mu\overset{T\rightarrow\infty}{\longrightarrow} 1$. If $\chi_R^\mu$ decreases as the temperature is lowered during the renormalization group flow the system neglects the perturbation. On the other hand, any significant increase $\chi_R^\mu> 1$ can be interpreted as an indication of nematic order corresponding to the perturbation imposed.

As explained, the magnitudes of the response functions $\chi_R^\mu$ allow one to estimate the tendency of a system to realize different lattice nematic orders. However, a comment of caution is necessary, because this approach cannot be used to detect actual nematic long range order or scaling collapses at the corresponding phase transitions [as described in Eq.~(\ref{eq:corrlength}) for magnetic long-range order]. This is because nematic correlations that become long-range at such transitions are described by four-spin correlators $\sim\langle S_k^\mu S_l^\mu S_m^\mu S_n^\mu\rangle$ which, in terms of Majorana vertex functions, correspond to four-particle vertices. However, within the current truncation of PMFRG equations, the renormalization group flows of vertices higher than the two-particle vertex are neglected such that nematic correlations and nematic phase transitions are not directly accessible. Keeping such higher vertex functions in the renormalization group flow would be an extraordinarily difficult method extension that exceeds available numerical resources. The response functions $\chi_R^\mu$ considered here bypass these difficulties but are subject to the aforementioned limitations. In Sec.~\ref{sec:nematic}, we compare the relative strengths of responses $\chi_R^\mu$ in different parts of the phase diagram and for different perturbation patterns in order to identify the dominant responses, but without being able to rigorously link a large susceptibility $\chi_R^\mu>1$ to actual nematic long-range order. Despite these limitations, nematic response functions from PFFRG and PMFRG have been very successfully used for characterizing magnetically disordered phases in past applications~\cite{PhysRevB.81.144410,PhysRevB.105.054426,PhysRevB.106.235137,PhysRevB.104.L220408,PhysRevB.105.L041115,PhysRevX.9.011005,PhysRevB.94.140408,PhysRevB.94.224403}.

The numerical specifications of our algorithm for solving the PMFRG equations are as follows. For the computation of magnetic phase boundaries we use 48 positive Matsubara frequencies and system sizes of 459, 1029 and 1941 correlated spins. Critical temperatures are determined from a collapse of $\xi^\mu/L$ of the two largest simulated systems. We start the renormalization group flow at $T=10^{5}J$ and evaluate observables at approximately 400 temperatures and 90 values for $\theta$. To solve the differential equation, we use the Julia package ``OrdinaryDiffEq'' with the solving method DP5 and an accuracy of $10^{-7}$.

\subsection{Classical Monte Carlo}

The classical Monte Carlo calculations for the XXZ Hamiltonian in Eq.~(\ref{eq:XXZhamiltonian}) are performed by treating spins as three-component unit vectors, $\boldsymbol{S} = (S^x,S^y,S^z)$ where $|\boldsymbol{S}| = 1$. This means that even in the XY or Ising limits we have extra components which do not contribute to the energy but can fluctuate at finite temperatures. The extra freedom to fluctuate causes the critical temperature to be lowered compared to the pure XY or Ising cases with two or one-component spins, respectively.

Our calculations are performed on cubic lattices of $N=n_u L^3$ sites, where $L$ is the number of cubic unit cells in each Cartesian direction and $n_u$ is the number of spins in the cubic unit cell, $n_u = 16$. The results shown in Fig.~\ref{fig:phasediagram}(a) correspond to $L=12$, implying $N=27648$ spins. The energy and specific heat are obtained through a logarithmic cool-down protocol starting at $T/J = 4$ and finishing at $T/J=0.01$ with 200 steps. At each temperature step, $10^5$ Monte Carlo steps are performed. The first half is used for thermalization, while the second half for collecting measurements. Each Monte Carlo step consists of $N$ single spin-update trials and $2N$ over-relaxation steps. For the spin-update trials we use the adaptive Gaussian step to ensure keeping a $50\%$ acceptance rate~\cite{Alzate19a}. 

\section{Results} \label{sec:results}
Our main results from PMFRG are summarized in Fig.~\ref{fig:phasediagram}, where the $\theta$-$T$ phase diagram is presented in Fig.~\ref{fig:phasediagram}(a), the spin structure factors in various different regions are shown in Fig.~\ref{fig:phasediagram}(b) and nematic responses are illustrated in Fig.~\ref{fig:phasediagram}(c). In the following three Subsections~\ref{sec:phase_diagram}, \ref{sec:ssf}, and \ref{sec:nematic} we discuss the physical details contained in Fig.~\ref{fig:phasediagram}(a), (b), (c), respectively.

\begin{figure}
		\includegraphics[width=0.49\textwidth]{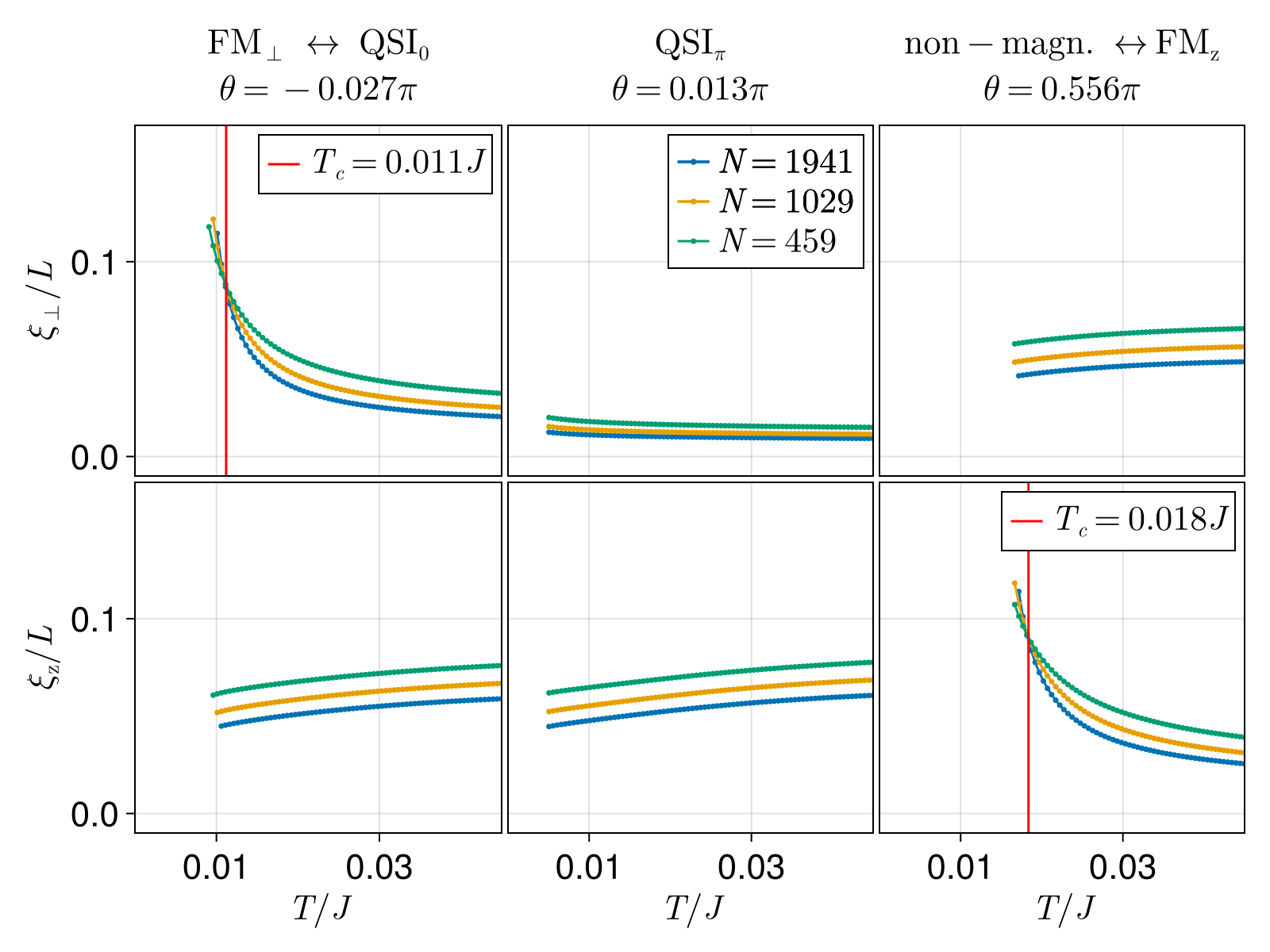}
    \vspace{-15pt}
	\caption[Correlation Lengths]{Correlation lengths $\xi_\mu/L$ as a function of temperature $T$ in the transverse ($\mu = \perp$) and longitudinal $(\mu=z)$ directions for different $\theta$. Green, yellow, and blue curves correspond to $N=459,1029,1941$ sites, respectively. Red vertical lines mark phase transitions indicated by a scaling collapse.}
	\label{fig:corr_lengths}
\end{figure}

\subsection{Magnetic phase diagram}\label{sec:phase_diagram}
The phase diagram in Fig.~\ref{fig:phasediagram}(a) shows two large domes of magnetic long-range order, corresponding to ferromagnetic order in the transverse $x$-$y$ plane (denoted FM$_\perp$) and Ising-type ferromagnetic order along the longitudinal $z$-axis (denoted FM$_z$). The critical temperatures in units of $J$ reach their maxima in both domes at the ferromagnetic XY model ($\theta=3\pi/2$) and at the ferromagnetic Ising model ($\theta=\pi$), respectively. A kink-like minimum of the transition temperature is observed at the ferromagnetic Heisenberg point ($\theta=5\pi/4$) where both magnetically ordered phases meet. For the ferromagnetic Ising model we can compare our critical temperature $T_c=1.18J$ with the classical Monte-Carlo result from Ref.~\cite{SOLDATOV2017707} (see Tab.~\ref{tab:Tcs}) which finds $T_c=1.053J$ when using the Ising spin normalization $S_i^z=\pm1/2$. With a deviation of $\sim12\%$ this value is in good agreement with our PMFRG result. It is worth noting that a classical Ising model generally poses the same methodological challenges to the PMFRG as a quantum model. A somewhat reduced accuracy ($\sim22\%$, see Tab.~\ref{tab:Tcs}) of our PMFRG transition temperature is observed for the ferromagnetic Heisenberg model, where we can compare with the quantum Monte-Carlo result of Ref.~\cite{PhysRevB.96.174419}.

\setlength{\tabcolsep}{4pt}
\renewcommand{\arraystretch}{1.4}
\begin{table}[t]
\centering
    \begin{tabular}{| l | c | c | c | c |}
        \hline
        \multicolumn{5}{|c|}{Critical temperatures $T_c/J$} \\ \hline\hline
        \multicolumn{2}{|c|}{}  & FM Ising & FM Heisenberg & FM XY \\ \hline
        PMFRG & quantum & 1.18 & 0.62 & 1.00 \\ \hline
        \multirow{3}{4em}{MC} & quantum & - & 0.508~\cite{PhysRevB.96.174419} & -\\ \cline{2-5}
                & classical & 1.57(3) & 0.92(2) & 1.44(2) \\ 
                &  & 4.21~\cite{SOLDATOV2017707} &  &  \\\hline
    \end{tabular}
    \caption[Table critical temperatures]{Critical temperatures in comparison between PMFRG and Monte Carlo (MC) methods for both quantum and classical systems. Numbers without a citation are from this work. We assume classical spins to be three-component vectors $\boldsymbol{S}_i=(S_i^x,S_i^y,S_i^z)$ with $|\boldsymbol{S}_i|=1$. In contrast, the classical result from Ref.~\cite{SOLDATOV2017707} is obtained by using {\it discrete} Ising spins $S_i^z=\pm1$. No literature results are available for the quantum ferromagnetic XY model.}
    \label{tab:Tcs}
\end{table}

\setlength{\tabcolsep}{12pt}
\renewcommand{\arraystretch}{1.4}
\begin{table}[t]
\centering
    \begin{tabular}{| l | c | c |}
        \hline
        \multicolumn{3}{|c|}{Critical couplings $\theta/\pi$} \\ \hline\hline
                & FM$_\perp$ $\leftrightarrow$ QSI$_0$ & non-magn. $\leftrightarrow$ FM$_z$ \\ \hline
       PMFRG    & -0.027 & 0.554 \\ \hline
       QMC      & -0.033~\cite{PhysRevLett.100.047208} & -\\ \hline
       CMFT     & -0.081~\cite{PhysRevLett.121.067201} & 0.613~\cite{PhysRevLett.121.067201} \\ \hline
       CMC       & -0.10(1) & 0.60(1) \\ \hline
    \end{tabular}
    \caption[Table different $\theta$]{Critical couplings $\theta/\pi$ of the phase transitions between the FM$_\perp$ and QSI$_0$ phases (middle column) and between the non-magnetic and FM$_z$ phases (right column) from PMFRG and classical Monte Carlo (CMC), as obtained in this work, as well as from quantum Monte Carlo (QMC)~\cite{PhysRevLett.100.047208} and cluster-mean field theory (CMFT)~\cite{PhysRevLett.121.067201}.}
    \label{tab:thetas}
\end{table}

Between the two magnetically ordered regimes we find a large region $\theta\in[-0.027\pi, 0.554\pi]$ where no magnetic phase transition occurs down to the lowest accessible temperatures. This low-temperature non-magnetic regime includes the antiferromagnetic Ising ($\theta=0$), Heisenberg ($\theta=\pi/4$) and XY ($\theta=\pi/2$) models. Near the boundaries of this region at $\theta\lesssim-0.027\pi$ and $\theta\gtrsim0.554\pi$ the transition temperatures can be traced down to very low temperature. As illustrated in Fig.~\ref{fig:corr_lengths}, even at temperatures $T$ two orders of magnitude smaller than $J$ a clear scaling collapse is observed. The ability of the $T$-flow approach to reach low temperatures (significantly lower than for $\Lambda$-flow schemes) has already been observed and explained in Ref.~\cite{PhysRevB.109.195109}. Interestingly, here we detect magnetic phase transitions at even smaller temperatures (almost one order of magnitude) than in Ref.~\cite{PhysRevB.109.195109}. The zero-temperature transition between the QSI$_0$ and FM$_\perp$ phases has previously been studied in detail since this region is accessible by sign-problem free quantum Monte-Carlo. The location of the phase transition from these works, $\theta=-0.033\pi$~\cite{PhysRevLett.115.077202,PhysRevLett.100.047208}, agrees well with our prediction $\theta=-0.027\pi$ indicating a robust functioning of the PMFRG even in the low temperature regime $T\ll J$. This is remarkable because the PMFRG is perturbatively controlled only at $T\gg J$~\cite{PhysRevB.103.104431} which highlights the capabilities of the approach beyond perturbation theory. On the other hand, the upper limit of the low-temperature non-magnetic phase from PMFRG at $\theta=0.554\pi$ is significantly lower than in previous studies of the quantum model ($\theta=0.613\pi$ in Ref.~\cite{PhysRevLett.121.067201}), see Tab.~\ref{tab:thetas}, indicating a larger stability of the FM$_z$ phase within PMFRG.

It is instructive to compare these findings with results from the classical XXZ pyrochlore model to estimate the impact of quantum fluctuations. For better comparison, we performed our own classical Monte Carlo simulations for three-component spins $\boldsymbol{S}_i=(S_i^x,S_i^y,S_i^z)$ normalized as $|\boldsymbol{S}_i|=1$. The magnetic phase transitions from these calculations are also shown in Fig.~\ref{fig:phasediagram}(a). Overall, the classical calculations reveal a similar two-dome structure than the quantum PMFRG results. Some caution is required when comparing explicit numbers for the transition temperatures in the classical and quantum cases since even at the ferromagnetic Ising model different transition temperatures are expected in both cases. This is because within our classical Monte-Carlo calculations we simulate the ferromagnetic Ising model with continuous three-component spins instead of discrete Ising variables $S_i^z=\pm1/2$ to correctly connect to the XXZ model at finite $J^\perp\neq0$ (this means that in our classical simulations $S_i^x$ and $S_i^y$ can take finite values, but these components do not contribute to the energy). As a result of the additional possibilities to fluctuate into transverse spin components our classical simulations show a significantly reduced ordering temperature at the ferromagnetic Ising limit $\theta=\pi$ compared to computations with discrete Ising variables, see Tab.~\ref{tab:Tcs}. On the other hand, a direct comparison between classical and quantum results can be made for the $\theta$-extent of the low-temperature non-magnetic phase which is found at $\theta\in[-0.10\pi,0.60\pi]$ in our classical Monte-Carlo simulations. The comparison with PMFRG indicates a suppression of the non-magnetic phase and an increase of the ordered regimes by quantum fluctuations at {\it both} transitions. At the lower transition into the FM$_\perp$ phase, this quantum order-by-disorder effect is known from previous quantum Monte-Carlo results~\cite{PhysRevLett.115.077202,PhysRevLett.100.047208}. In addition, here we find a similar quantum order-by-disorder mechanism at the upper transition into the FM$_z$ phase which has not yet been observed. In fact, previous quantum investigations have found the opposite, i.e., a decrease of the ordered phase by quantum fluctuations in this parameter region~\cite{benton2020,PhysRevLett.121.067201}.

\begin{figure*}
		\includegraphics[width=1.0\textwidth]{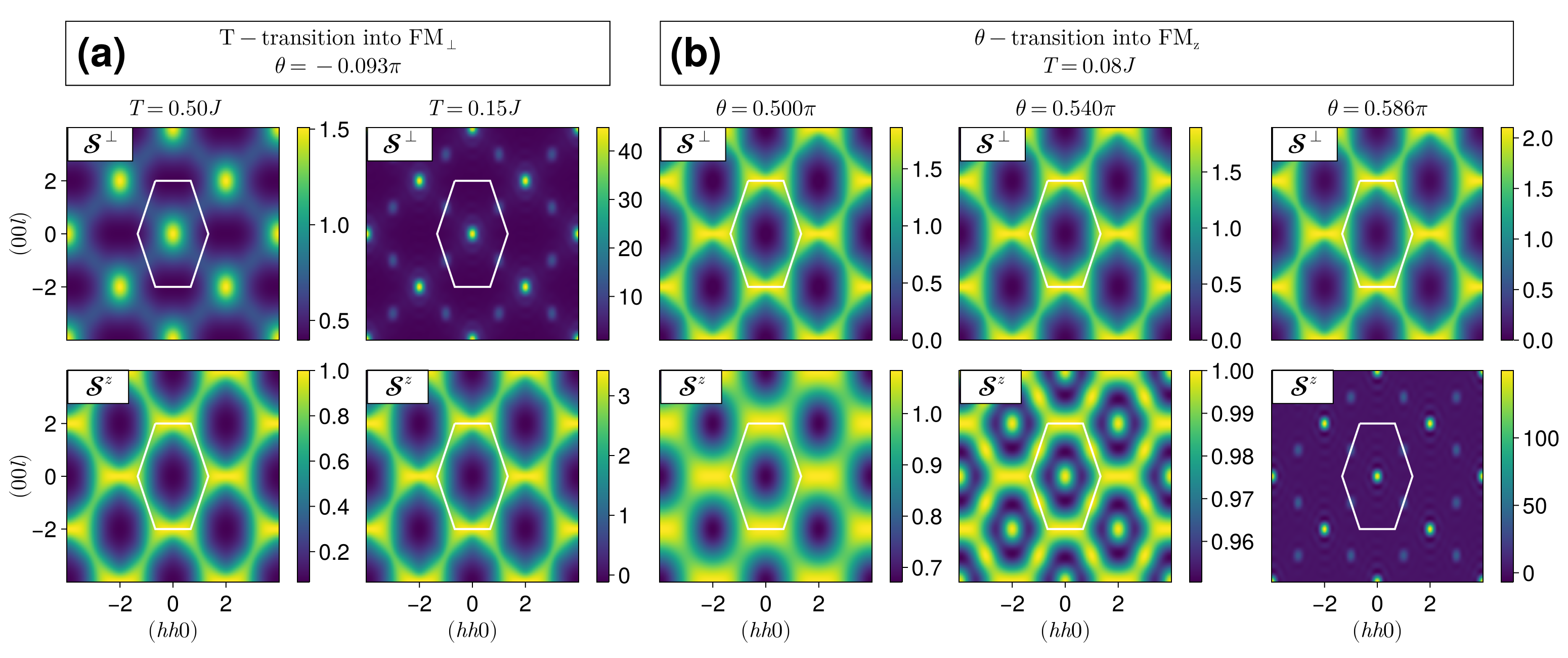}
    \vspace{-15pt}
	\caption[Additional Structure Factors]{(a) Spin structure factors $\mathcal{S}^\perp(\boldsymbol{Q})$ and $\mathcal{S}^z(\boldsymbol{Q})$ in the $(hhl)$ plane at constant $\theta = -0.093\pi$ (corresponding to $J^\perp/J^z \approx -0.3$) for two temperatures $T=0.5J$ and $T=0.15J$, the latter being right at the phase transition into the FM$_\perp$ order. (b) Spin structure factors $\mathcal{S}^\perp(\boldsymbol{Q})$ and $\mathcal{S}^z(\boldsymbol{Q})$ in the $(hhl)$ plane at constant temperature $T=0.08J$, varying $\theta$ from the antiferromagnetic XY model ($\theta=\pi/2$) into the FM$_z$ phase.}
	\label{fig:add_SF}
\end{figure*}

\subsection{Spin structure factors}\label{sec:ssf}
Further insights into the spin correlations in the different magnetically ordered and disordered phases are obtained by investigating the spin structure factor. Firstly, ferromagnetic long-range order in a spin component $\mu$ is indicated by a pattern of sharp magnetic Bragg peaks in $\mathcal{S}^{\mu}(\boldsymbol{Q})$ as shown in Fig.~\ref{fig:phasediagram}(b), right panel (where an exemplary plot for the ferromagnetic Heisenberg model is presented) with a dominant ordering peak at $\boldsymbol{Q}=(0,0,0)$ and smaller peaks at $\boldsymbol{Q}=2\pi(\pm1,\pm1,\pm1)$ [i.e., $(h,k,l)=(\pm1,\pm1,\pm1)$]. In contrast, in non-magnetic phases the signal is more spread over the Brillouin zone. Particularly, for the antiferromagnetic Ising model at $\theta=0$ the longitudinal spin structure factor $\mathcal{S}^{z}(\boldsymbol{Q})$ [see Fig.~\ref{fig:phasediagram}(b)] shows the characteristic pattern known from classical spin ice, with pinch points located at $\boldsymbol{Q}=2\pi(0,0,2)$ (and symmetry-related positions)~\cite{PhysRevLett.93.167204}. In the exact spin structure factor of classical spin ice the pinch points are sharp and singular, which is a consequence of the dipolar correlations induced by the spin ice constraint $S_{t,1}^z+S_{t,2}^z+S_{t,3}^z+S_{t,4}^z=0$ for the four spins in each tetrahedron $t$. In our PMFRG result, however, the pinch points show a small broadening, which is an artifact of the approximations involved, especially the truncation of renormalization group equations. While directly evident on the Hamiltonian level, the spin ice constraints are highly non-trivial in terms of the renormalization group flow of Majorana vertex functions and cannot be expected to be exactly fulfilled at any level of truncation. The transverse spin structure factor $\mathcal{S}^{xx}(\boldsymbol{Q})=\mathcal{S}^{yy}(\boldsymbol{Q})\equiv\mathcal{S}^{\perp}(\boldsymbol{Q})$ is completely flat at the Ising point. This flat signal originates from the onsite spin correlator $\langle S_i^x S_i^x\rangle=\langle S_i^y S_i^y\rangle=1/4$ which is exactly captured within PMFRG~\cite{PhysRevB.106.235113}.

Moving slightly away from the antiferromagnetic Ising point, the longitudinal spin structure factor $\mathcal{S}^{z}(\boldsymbol{Q})$ first remains unchanged, however, the transverse components $\mathcal{S}^{\perp}(\boldsymbol{Q})$ develop a small modulation in reciprocal space, see Fig.~\ref{fig:phasediagram}(b). This modulation describes quantum fluctuations in the space of classical spin ice states and, hence, directly characterizes the U(1) spin liquid in this quantum spin ice regime. At $\theta=-0.013\pi$ in the QSI$_0$ phase, $\mathcal{S}^{\perp}(\boldsymbol{Q})$ shows faint peaks at the Brillouin zone center $\boldsymbol{Q}=(0,0,0)$ while at $\theta=0.013\pi$ in the QSI$_\pi$ phase small maxima are observed at $\boldsymbol{Q}=2\pi(0,0,2)$. In the latter case the signal resembles a smeared version of the longitudinal spin structure factor $\mathcal{S}^{z}(\boldsymbol{Q})$. These patterns agree nicely with the predictions from gauge mean-field theory~\cite{PhysRevLett.132.066502}. Particularly, the maxima at $\boldsymbol{Q}=(0,0,0)$ and $\boldsymbol{Q}=2\pi(0,0,2)$ in the transverse spin structure factor $\mathcal{S}^{\perp}(\boldsymbol{Q})$, representing a characteristic distinguishing feature between QSI$_0$ and QSI$_\pi$, respectively~\cite{PhysRevLett.132.066502} are correctly captured by PMFRG. Increasing $\theta$ towards the Heisenberg point $\theta=\pi/4$ the longitudinal spin structure factor $\mathcal{S}^{z}(\boldsymbol{Q})$ remains qualitatively unchanged (not shown), while the transverse spin structure factor increases until at $\theta=\pi/4$ all components of $\mathcal{S}^{\mu}(\boldsymbol{Q})$ become identical.

In Fig.~\ref{fig:add_SF} we show additional spin structure factors to discuss specific behaviors of $\mathcal{S}^{\mu}(\boldsymbol{Q})$ near magnetic phase transitions in more detail [the angles $\theta$ and temperatures $T$ of these plots are marked in Fig.~\ref{fig:phasediagram}(a) by red crosses]. A typical cooling process from the non-magnetic phase into the low-temperature FM$_\perp$ phase at constant $\theta$ is presented in Fig.~\ref{fig:add_SF}(a) for the case $\theta=-0.093\pi$. At $T=0.5J$, well inside the non-magnetic phase, $\mathcal{S}^{\perp}(\boldsymbol{Q})$ shows a square-shaped network of connected peaks at $\boldsymbol{Q}=(0,0,0)$ and at the centers of the other Brillouin zones. The observed pattern resembles the one in $\mathcal{S}^{\perp}(\boldsymbol{Q})$ of the QSI$_0$ phase in Fig.~\ref{fig:phasediagram}(b), but with stronger modulations. This demonstrates that effective quantum spin liquid behavior can also emerge at temperatures {\it above} a magnetically ordered phase. However, we are unable to quantify the extent of the quantum spin liquid phase in $T$ and $\theta$ due to continuously varying spin structure factors. Cooling the system to $T=0.15J$, which corresponds to the phase transition, the peaks at $\boldsymbol{Q}=(0,0,0)$ in $\mathcal{S}^{\perp}(\boldsymbol{Q})$ become sharper and the signal of the square-shaped network concentrates at the subdominant peaks at $\boldsymbol{Q}=2\pi(\pm1,\pm1,\pm1)$. We note that the finite height of the magnetic Bragg peaks at the phase transition is a consequence of the finite system size in our PMFRG calculations.

In Fig.~\ref{fig:add_SF}(b) we show spin structure factors for the system's transition from the magnetically disordered antiferromagnetic XY model into the ordered FM$_z$ phase as $\theta$ is increased and $T$ is kept small and constant at $T=0.08J$. At the antiferromagnetic XY point $\theta=\pi/2$ the transverse spin structure factor $\mathcal{S}^{\perp}(\boldsymbol{Q})$ shows a typical signal of broadened pinch points that closely resembles the one in $\mathcal{S}^{z}(\boldsymbol{Q})$ of the Ising model. Here, however, the broadening which indicates a (partial) violation of the spin ice constraint $\langle S_{t,1}^x+S_{t,2}^x+S_{t,3}^x+S_{t,4}^x\rangle=\langle S_{t,1}^y+S_{t,2}^y+S_{t,3}^y+S_{t,4}^y\rangle\neq0$ is a physical feature and no numerical artifact. This is because $S_{t,1}^x+S_{t,2}^x+S_{t,3}^x+S_{t,4}^x$ and $S_{t,1}^y+S_{t,2}^y+S_{t,3}^y+S_{t,4}^y$ are non-commuting operators for two overlapping or identical tetrahedra, preventing the simultaneous fulfillment of all spin ice constraints. Interestingly, even though $J^z$ vanishes in the XY model, the signal from $\mathcal{S}^{\perp}(\boldsymbol{Q})$ leaks into the longitudinal spin structure factor $\mathcal{S}^{z}(\boldsymbol{Q})$ but the pattern in $\mathcal{S}^{z}(\boldsymbol{Q})$ is somewhat smeared and reduced compared to $\mathcal{S}^{\perp}(\boldsymbol{Q})$. Upon increasing $\theta$ away from the XY point we find a $\theta$ regime where $\mathcal{S}^{z}(\boldsymbol{Q})$ shows both, the pattern inherited from the antiferromagnetic transverse couplings $J^\perp>0$ as well as peaks at the Brillouin zone centers due to $J^z<0$ indicating the proximity to the FM$_z$ phase, see middle panel of Fig.~\ref{fig:add_SF}(b) at $\theta=0.54\pi$. Upon further increasing $\theta$ the system develops FM$_z$ order as indicated by the peak structure of $\mathcal{S}^{z}(\boldsymbol{Q})$ in the right panel of Fig.~\ref{fig:add_SF}(b). Interestingly, during the system's evolution as a function of $\theta$ as shown in Fig.~\ref{fig:add_SF}(b), the transverse spin structure factor $\mathcal{S}^{\perp}(\boldsymbol{Q})$ remains largely unchanged.

\begin{figure}
		\includegraphics[width=0.49\textwidth]{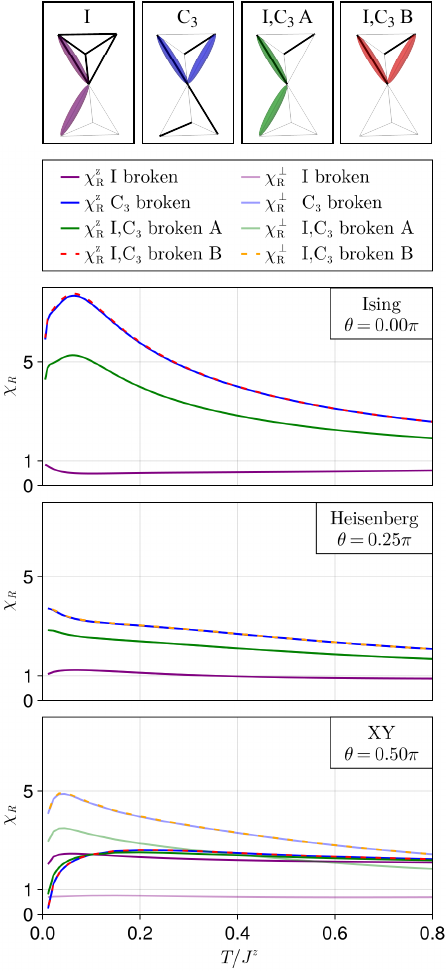}
    \vspace{-15pt}
	\caption[Dimer Responses]{
    Responses to lattice symmetry breaking perturbations. The top panel shows the perturbation patters of strengthened (thick black) and weakened (thin gray) bonds. The colored bonds indicate the strong and weak bonds that define the response function $\chi_R^\mu$ in Eq.~(\ref{eq:response}). The three plots show the temperature dependence of the transverse $\chi_R^\perp$ and longitudinal $\chi_R^z$ responses for all considered perturbations for the Ising, Heisenberg and XY models.}
	\label{fig:dimer_response}
\end{figure}

\subsection{Nematic responses}\label{sec:nematic}
To obtain more insights into the system's properties in magnetically disordered phases, we compute responses to lattice symmetry breaking perturbation as described in Sec.~\ref{sec:pmfrg}. To this end, we strengthen and weaken the couplings according to Eq.~(\ref{eq:couplings2}) with a small parameter $\delta=0.01$ and calculate $\chi_R^\mu$ defined in Eq.~(\ref{eq:response}) as a function of temperature. The patterns of strengthened and weakened couplings are depicted as thick black and thin gray bonds in the top panel of Fig.~\ref{fig:dimer_response}. These perturbations are the same as those considered in Refs.~\cite{PhysRevB.105.054426,PhysRevB.106.235137}. The first pattern strengthens the couplings on one tetrahedron type (e.g. the ``up-tetrahedron'') and weakens them on the other, leading to inversion (``I'') symmetry breaking about pyrochlore sites. The second pattern strengthens bonds along straight chains running through the entire system. Note that there are three types of such chains in the pyrochlore lattice along different directions, which do not cross. This perturbation breaks the three-fold rotation symmetry C$_3$ about an axis that connects the centers of adjacent tetrahedra. The third pattern describes a dimer covering on the pyrochlore lattice where two opposite bonds on one tetrahedron type are strengthened while all other bonds (including those of the other tetrahedron type) are weakened. This perturbation breaks both inversion and three-fold rotation symmetry (``I, C$_3$''). Furthermore, the colored bonds in the top panel of Fig.~\ref{fig:dimer_response} -- in each case one strengthened and one weakened bond -- mark the correlations $\langle S_k^\mu S_l^\mu\rangle$ and $\langle S_m^\mu S_n^\mu\rangle$ that are used to calculate the response function $\chi_R^\mu$ in Eq.~(\ref{eq:response}). For symmetry reasons, for the first two perturbations all possibilities for choosing strengthened bonds $(k,l)$ and weakened bonds $(n,m)$ in Eq.~(\ref{eq:response}) lead to the same response function $\chi_R^\mu$. However, for the third perturbation pattern, two symmetry inequivalent choices for the bonds $(k,l)$ and $(n,m)$ exist, leading to different response functions denoted ``A'' and ``B''.

The results for $\chi_R^\mu$ from PMFRG for all considered perturbations are presented in the three plots in Fig.~\ref{fig:dimer_response} showing the Ising, Heisenberg and XY cases. We distinguish between responses $\chi_R^\perp$ ($\chi_R^z$) probing transverse (longitudinal) spin correlations, see Eq.~(\ref{eq:response}). In the isotropic Heisenberg case where $\chi_R^\perp=\chi_R^z$ the response to the C$_3$ symmetry breaking perturbation and the response B to the I and C$_3$ symmetry breaking perturbation are largest at all temperatures (the two responses are almost but not exactly identical). The susceptibility A for the latter perturbation is somewhat smaller and the response to only inversion symmetry breaking is weakest. These results can be compared with those from PFFRG in Ref.~\cite{PhysRevB.105.054426} where the same response functions for the pyrochlore Heisenberg model were calculated. This work also found an enhanced response to C$_3$ and combined I/C$_3$ symmetry breaking (but with a different overall magnitude of nematic susceptibilities) which was interpreted as a possible indication of lattice nematic order in this system. Our results for the smaller response functions are also qualitatively similar to Ref.~\cite{PhysRevB.105.054426}.

The responses to C$_3$ and combined I/C$_3$ symmetry breaking remain dominant in the Ising model (but show a characteristic decrease at small temperatures) and are larger than in the Heisenberg case. Note that the definition of $\chi_R^\mu$ in Eq.~(\ref{eq:response}) only allows the calculation of the longitudinal response $\chi_R^z$ in the Ising model. The transverse response $\chi_R^\perp$ is ill-defined in this case because all non-local $xx$ and $yy$ spin correlations vanish, $\langle S_i^x S_j^x\rangle=\langle S_i^y S_j^y\rangle=0$ for $i\neq j$. The XY model shows a qualitatively similar behavior to the Ising model: Responses to C$_3$ and combined I/C$_3$ breaking are dominant and enhanced compared to the Heisenberg model. In the XY model, however, the largest responses are along the transverse spin directions. In contrast to the Ising model where $\chi_R^\perp$ is ill-defined, in the XY model longitudinal susceptibilities $\chi_R^z$ can be computed, despite vanishing $J^z=0$, but they are smaller than $\chi_R^\perp$. The finite signal in $\chi_R^z$ for the XY model can be explained by the system's quantum nature, where the transverse response $\chi_R^\perp$ leaks into the longitudinal direction. A similar effect induces the finite spin structure factor $\mathcal{S}^z(\boldsymbol{Q})$ for the XY model in Fig.~\ref{fig:add_SF}(b).

Overall, these results show that the responses to C$_3$ and combined I/C$_3$ symmetry breaking are largest in all three models and increase further as one moves away from the Heisenberg case towards the Ising and XY models. To illustrate this behavior more clearly, in Fig.~\ref{fig:phasediagram}(a) and (c) we plot the dominant C$_3$ response throughout the non-magnetic phase as a function of $\theta$ and $T$. One observes a pronounced increase of $\chi_R^\mu$ on both sides of the Heisenberg point $\theta=\pi/4$ where for $\theta<\pi/4$ ($\theta>\pi/4$) the longitudinal response $\chi_R^z$ (transverse response $\chi_R^\perp$) is largest. The maxima of $\chi_R^\mu$ as a function of $\theta$ are approximately assumed for the Ising and XY models.

It is generally expected that C$_3$ symmetry breaking in the Ising model will give rise to a large response at low temperatures. This is because of the classical nature of this model where for any finite $\delta>0$, the perturbation selects a subset of the classical spin ice states (with purely antiferromagnetic spin configurations along the three chains). Thus, even for infinitesimally small $\delta$ the difference $\langle S_k^\mu S_l^\mu\rangle-\langle S_m^\mu S_n^\mu\rangle$ between spin correlations on strengthened and weakened bonds is of order unity and $\chi_R^z\sim1/\delta$ [see Eq.~(\ref{eq:response})]. It follows that $\chi_R^z$ must diverge for $\theta\rightarrow0$ and $T\rightarrow0$ in the limit of small $\delta$. As explained, in PMFRG we find this expected strong increase of $\chi_R^z$ near the Ising model at low temperatures. The fact that our responses do not fully diverge but remain finite in this limit is probably an artifact from the truncation of the renormalization group equations. In other words, large responses to symmetry breaking perturbations near the Ising limit should not be interpreted as nematic order but rather indicate that the system becomes classical in this part of the phase diagram. However, this explanation does not apply to the strong responses we observe near the XY model, since this system is a true quantum model. We can therefore conclude that the pyrochlore XY model shows strong tendencies for realizing a nematic ground state with broken C$_3$ symmetry or a combination of broken C$_3$ and inversion symmetries. Remarkably, these tendencies are even more pronounced than in the pyrochlore Heisenberg model for which there is compelling evidence for a lattice symmetry broken ground state from several different numerical approaches~\cite{PhysRevLett.126.117204,PhysRevX.11.041021,PhysRevB.105.054426,PhysRevLett.131.096702,pohle2023groundstates12pyrochlore}. Due to the continuously varying responses $\chi_R^\perp$ near the XY model and the general limitations of our approach to probe long-range nematic correlations (see discussion in Sec.~\ref{sec:pmfrg}) we are not able to identify the boundaries of this potential lattice symmetry broken phase and how it connects to the QSI$_\pi$ phase.

\section{Conclusion} \label{sec:summary}
We use the recently developed PMFRG in the $T$-flow formalism~\cite{PhysRevB.109.195109} to calculate the phase diagram of the spin-$1/2$ nearest neighbor XXZ pyrochlore model in the full parameter range $\theta\in[0,2\pi]$ of ferromagnetic and antiferromagnetic couplings and for temperatures down to $T\approx0.01J$. Even though the method is expected to lose quantitative accuracy at $T<J$, our phase transition at lowest temperatures between the QSI$_0$ and FM$_\perp$ regimes is in very good agreement with quantum Monte Carlo. Particularly, even at $T\approx0.01J$ magnetic phase transitions are accurately indicated by a critical scaling of the correlation length, demonstrating that the PMFRG is applicable at temperatures well beyond the regime where the method is perturbatively controlled. While quantum order-by-disorder has previously been found near the transition between the QSI$_0$ and FM$_\perp$ phases, here we identify a similar effect near the upper limit $\theta=0.554\pi$ of our low-temperature non-magnetic phase, where classical Monte Carlo finds the transition into the FM$_z$ phase at larger $\theta=0.60\pi$. This is in contrast to a previous investigation of the quantum model using variational and exact diagonalization approaches~\cite{PhysRevLett.121.067201} where quantum fluctuations were found to increase the non-magnetic phase. We further discuss spin structure factors in various parts of the phase diagram. While deep in the magnetically disordered phase the spin structure factor shows typical patterns of broadened pinch points, near magnetic phase boundaries additional features due to the proximity to long-range order are observed.

A last focus of our work is on responses to lattice symmetry breaking perturbations. We find the responses to C$_3$ and combined I/C$_3$ symmetry breaking enhanced at the antiferromagnetic Heisenberg point at low temperatures, in agreement with previous PFFRG studies~\cite{PhysRevB.105.054426}. Interestingly, these responses increase even further when moving from the Heisenberg to the XY model. We therefore propose a lattice nematic state as a promising candidate for the ground state of the antiferromagnetic XY model.

A previous work identified a spin-nematic ground state in the antiferromagnetic XY model~\cite{PhysRevLett.121.067201}. This is an interesting alternative possibility as it resembles the behavior of the corresponding classical model, where thermal order-by-disorder selects collinear spin configurations~\cite{PhysRevB.58.12049,PhysRevLett.80.2929,PhysRevX.7.041057}. Within PMFRG, such in-plane spin nematic order can in principle be probed by detuning the transverse couplings $J^x$ and $J^y$ homogeneously on all nearest neighbor lattice bonds and calculating the system's response to a small perturbation $\delta=J^x-J^y$. The current implementation of the PMFRG, however, does not allow for such a modification because spin rotation symmetry around the $z$ axis is explicitly assumed in our PMFRG equations. An extension of the PMFRG to general anisotropic spin interactions will require a re-derivation of the renormalization group equations giving rise to more complicated equations. Although this is expected to significantly increase the numerical runtimes, such a method extension still seems feasible, but we leave it for future work. The resulting approach will allow the calculation of spin-nematic responses and the investigation of larger classes of spin Hamiltonians such as XYZ models~\cite{PhysRevB.109.184421}. This opens up interesting research perspectives, e.g. comparisons with experimental results for dipolar-octopolar spin ice materials where such XYZ couplings are realized~\cite{PhysRevX.12.021015,bhardwaj2022,PhysRevX.14.011005,sibille_quantum_2020,gao_experimental_2019,gao2024}.

\section{Acknowledgement}
We would like to thank Nils Niggemann, Daniel Lozano-G\'omez, Kristian Tyn Kai Chung and Michel Gingras for fruitful discussions. We acknowledge support from the Deutsche Forschungsgemeinschaft (DFG, German Research Foundation), within Project-ID 277101999 CRC 183 (Project
A04). We acknowledge the use of the JUWELS cluster at the Forschungszentrum J\"ulich and the HPC Service of ZEDAT, Freie Universit\"at Berlin.

\appendix*

\section{Details of the PMFRG} \label{sec:PMFRG_app}
The derivation of the PMFRG equations make systematic use of  symmetries to obtain simple dependencies of fermionic propagators $G^{ij}_{\mu\nu}(\omega_1,\omega_2)$ and vertex functions $\Gamma^{ijkl}_{\mu\nu\alpha\beta}(\omega_1,\omega_2,\omega_3,\omega_4)$ in their arguments, such that the resulting flow equations can be tackled numerically. Here, $i,j,k,l$ are site indices, $\mu,\nu,\alpha,\beta\in\{x,y,z\}$ are Majorana flavors and $\omega$ are fermionic Matsubara frequencies. The one-particle Green's function $G_{12}$ and the connected two-particle Green's function $G^c_{1234}$ are defined as
\begin{align}
    G_{12} &\equiv G^{ij}_{\mu\nu}(\omega_1,\omega_2) = \langle\eta_i^\mu(\omega_1)\eta_j^\nu(\omega_2)\rangle, \label{eq:propagator} \\
    G^c_{1234} &= - \hspace{-10pt}\int\displaylimits_{1'2'3'4'} \hspace{-10pt} G_{11'} G_{22'} G_{33'} G_{44'} \Gamma_{1'2'3'4'}, \label{eq:vertexfunction}
\end{align}
where the subscripts $1234$ are multi-indices including sites, Majorana flavors and Matsubara frequencies, such that any object $g$ can be briefly written as $g_1 \equiv g^i_\mu(\omega_1)$.

In the following, we state the consequences of all symmetries (as well as $\mathbb{Z}_2$ gauge freedom under $\eta\rightarrow -\eta$, Hermiticity and anti-commutation relations of Majorana operators) on the argument structure of one- and two-particle Green's functions. We use a lean notation in which we only show the indices that are affected by the symmetry transformation. Furthermore, $P(1234)$ denotes a permutation of the indices $1234$ and $\mathrm{sgn}(P)$ is $1$ ($-1$) for even (odd) permutations.\\

\noindent Anti-commutation relation of Majorana operators:
\begin{flalign}
    & \hspace{16pt} G_{12} = \langle\eta_1\eta_2\rangle = - \langle\eta_2\eta_1\rangle = -G_{21}, && \\
    & \hspace{16pt} G_{1234} = \mathrm{sgn}(P) \, G_{P(1234)}. \label{eq:G4permutation}&&
\end{flalign}
Hermiticity of Majorana operators:
\begin{flalign}
     & \hspace{16pt}  G_{12}^* = \langle(\eta_1\eta_2)^\dagger\rangle = \langle\eta_2\eta_1\rangle = G_{21}, && \\
     & \hspace{16pt}  G_{1234}^* = G_{1234}.&&
\end{flalign}
$\mathbb{Z}_2$ gauge redundancy:
\begin{flalign}
     & \hspace{16pt} G^{ij} = \delta_{ij} G^i, && \\
     & \hspace{16pt} G^{ijkl} = G^{ij} (\delta_{ik}\delta_{jl} + \delta_{ij}\delta_{kl} - \delta_{il}\delta_{jk}). &&
\end{flalign}
SO$(2)$ spin symmetry in the $x$-$y$ plane:
\begin{flalign}
     & \hspace{16pt} G_{\mu\nu} = \delta_{\mu\nu} G_{\mu\nu} = \delta_{\mu\nu} G_\mu, && \\
     & \hspace{16pt} G_{\mu\nu\alpha\beta} = G_{\mu\nu\alpha\beta}  (\delta_{\mu\nu}\delta_{\alpha\beta}  + \delta_{\mu\alpha}\delta_{\nu\beta} - \delta_{\mu\beta}\delta_{\nu\alpha}). &&
\end{flalign}
Time reversal symmetry:
\begin{flalign}
     & \hspace{16pt} G(\omega_1,\omega_2) = G(-\omega_1,-\omega_2)^*, && \\
     & \hspace{16pt} G(\omega_1,\omega_2,\omega_3,\omega_4) = G(-\omega_1,-\omega_2,-\omega_3,-\omega_4). &&
\end{flalign}
Time translation symmetry:
\begin{flalign}
    & \hspace{16pt} G(\omega_1,\omega_2) = \beta \delta_{\omega_1, -\omega_2} G(\omega_2), && \\
    & \hspace{16pt} G(\omega_1,\omega_2,\omega_3,\omega_4) = G(s,t,u) \delta_{\omega_1+\omega_2+\omega_3+\omega_4,0}, && \label{eq:G4Ttranslation}
\end{flalign}
where $\beta$ is the inverse temperature. In the last equation we make use of the following bosonic transfer frequencies,
\begin{align}
    s = \omega_1 + \omega_2 = -\omega_3 -\omega_4, \\
    t = \omega_1 + \omega_3 = -\omega_2 -\omega_4, \\
    u = \omega_1 + \omega_4 = -\omega_2 -\omega_3.
\end{align}
The resulting parametrizations for the propagator $G$ and the two-particle vertex $\Gamma$ are
\begin{align}
G^{ij}_{\mu\nu}(\omega_1,\omega_2) = &\, G^i_\mu(\omega_2) \delta_{ij} \delta_{\mu\nu} \delta_{\omega_1,-\omega_2},\label{eq:proppara} \\
\Gamma^{ijkl}_{\mu\nu\alpha\beta}(\omega_1,\omega_2,\omega_3,\omega_4) = &\, \delta_{\omega_1+\omega_2+\omega_3+\omega_4,0} \notag \\
&\times\Big[\phantom{+}\Gamma_{\mu\nu\alpha\beta}^{ij}(s,t,u) \delta_{ik}\delta_{jl} \notag \\
&\phantom{\times\Big[} -\Gamma_{\mu\nu\alpha\beta}^{ik}(s,t,u) \delta_{il}\delta_{jk} \notag \\
&\phantom{\times\Big[} +\Gamma_{\mu\nu\alpha\beta}^{ik}(s,t,u) \delta_{ij}\delta_{kl} \Big]. \label{eq:vertexpara}
\end{align}
In the present case of a spin model with a SO(2) spin rotation symmetry in the $x$-$y$ plane, there are eight different flavor combinations for the vertex functions $\Gamma_{xxxx}$, $\Gamma_{zzzz}$, $\Gamma_{xxyy}$, $\Gamma_{xxzz}$, $\Gamma_{zzxx}$, $\Gamma_{xyxy}$, $\Gamma_{xzxz}$, $\Gamma_{zxzx}$. Although there are relations between some of these, we explicitly work with all eight vertex functions. This is because by using time reversal symmetry, vertex functions with negative Matsubara frequencies $s$, $t$, $u$ can always be mapped to positive ones, increasing numerical efficiency.

The resulting flow equations for the self-energy $\Sigma$ and the two-particle vertex $\Gamma$ for this parametrization read as
\begin{widetext}
\begin{align}
     &\frac{\mathrm{d}}{\mathrm{d}T} \Sigma^{T,j}_{\nu}(\omega_1) = \frac{1}{2} \sum_{\omega} \sum_i  \sum_{\mu} ~\Gamma^{T,ij}_{\mu\mu\nu\nu}(0, \omega-\omega_1,\omega+\omega_1) ~\frac{\mathrm{d}}{\mathrm{d}T} G^{T,i}_\mu(\omega), \label{eq:floweq1app} \\
       &\frac{\mathrm{d}}{\mathrm{d}T} \Gamma^{T,ij}_{\mu\nu\alpha\beta}(s,t,u) = X^{T,ij}_{\mu\nu\alpha\beta}(s,t,u) - \tilde{X}^{T,ij}_{\mu\alpha\nu\beta}(t,s,u) + \tilde{X}^{T,ij}_{\mu\beta\nu\alpha}(u,s,t), \phantom{\sum_\omega} \label{eq:floweq2app}\\
    &X^{T,ij}_{\mu\nu\alpha\beta}(s,t,u) = \frac{1}{2} \sum_{\lambda\rho} \sum_\omega \sum_k
    P^{T,kk}_{\lambda\rho}(\omega,\omega+s)
    \Gamma^{T,ik}_{\mu\nu\rho\lambda}(s,-\omega-\omega_2,\omega+\omega_1)
    \Gamma^{T,kj}_{\lambda\rho\alpha\beta}(s,-\omega+\omega_3,-\omega+\omega_4), \label{eq:bubble} \\
    &\tilde{X}^{T,ij}_{\mu\nu\alpha\beta}(s,t,u) = - \sum_{\lambda\rho} \sum_\omega
    P^{T,ij}_{\lambda\rho}(\omega,\omega+s)
    \Gamma^{T,ij}_{\mu\lambda\nu\rho}(\omega+\omega_1,s,-\omega-\omega_2)
    \Gamma^{T,ij}_{\lambda\alpha\rho\beta}(-\omega+\omega_3,s,-\omega+\omega_4).\label{eq:bubbletilde}
\end{align}
\end{widetext}
Here, we used the bubble propagator
\begin{equation}
P^{T,ij}_{\mu\nu}(\omega,\omega+s) = \frac{d}{dT}[G^{T,i}_\mu(\omega)G^{T,j}_\nu(\omega+s)]
\end{equation}
to simplify the notation~\cite{PhysRevB.109.195109}. The initial conditions for $\Sigma$ and $\Gamma$ at $T\rightarrow \infty$ are given by
\begin{align}
    & \Sigma^{\infty} = 0, \\
    & \Gamma^{\infty}_{xxxx} = \Gamma^{\infty}_{zzzz} = \Gamma^{\infty}_{xxyy} = \Gamma^{\infty}_{xxzz} = \Gamma^{\infty}_{zzxx} = 0, \\
    & \Gamma^{\infty, ij}_{xzxz} = \Gamma^{\infty, ij}_{zxzx} = -J^\perp_{ij}, \\
    & \Gamma^{\infty, ij}_{xyxy} = - J^z_{ij}.
\end{align}
From the solutions of the PMFRG equations in Eqs.~(\ref{eq:floweq1app}) and (\ref{eq:floweq2app}) we can calculate the equal-time spin-spin correlators $\langle S^\mu_i S_j^\mu\rangle=T\sum_\Omega \mathcal{C}_{ij}^{\mu}(\Omega)$ where the dynamical correlator $\mathcal{C}_{ij}^{\mu}(\Omega)$ with the bosonic Matsubara frequency $\Omega$ is given by
\begin{widetext}
\begin{align}
    &\mathcal{C}_{ij}^{x}(\Omega) = - \delta_{ij} \sum_{\omega_1} G_x^{T,i}(\omega_1)G_z^{T,i}(\omega_1+\Omega) \notag \\ 
    & \hspace{42pt} + \sum_{\omega_1 \omega_2} G_x^{T,i}(\omega_2+\Omega)G_z^{T,i}(\omega_2)G_x^{T,j}(\omega_1)G_z^{T,j}(\omega_1+\Omega) \Gamma_{xzxz}^{T,ij}(\Omega, \Omega+\omega_1+\omega_2, \omega_2-\omega_1), \label{eq:suscxx} \\
    &\mathcal{C}_{ij}^{z}(\Omega) = - \delta_{ij} \sum_{\omega_1} G_x^{T,i}(\omega_1)G_x^{T,i}(\omega_1+\Omega) \notag \\ 
    & \hspace{42pt} + \sum_{\omega_1 \omega_2} G_x^{T,i}(\omega_2+\Omega)G_x^{T,i}(\omega_2)G_x^{T,j}(\omega_1)G_x^{T,j}(\omega_1+\Omega) \Gamma_{xyxy}^{T,ij}(\Omega, \Omega+\omega_1+\omega_2, \omega_2-\omega_1). \label{eq:susczz}
\end{align}
\end{widetext}


\begin{thebibliography}{82}%
\makeatletter
\providecommand \@ifxundefined [1]{%
 \@ifx{#1\undefined}
}%
\providecommand \@ifnum [1]{%
 \ifnum #1\expandafter \@firstoftwo
 \else \expandafter \@secondoftwo
 \fi
}%
\providecommand \@ifx [1]{%
 \ifx #1\expandafter \@firstoftwo
 \else \expandafter \@secondoftwo
 \fi
}%
\providecommand \natexlab [1]{#1}%
\providecommand \enquote  [1]{``#1''}%
\providecommand \bibnamefont  [1]{#1}%
\providecommand \bibfnamefont [1]{#1}%
\providecommand \citenamefont [1]{#1}%
\providecommand \href@noop [0]{\@secondoftwo}%
\providecommand \href [0]{\begingroup \@sanitize@url \@href}%
\providecommand \@href[1]{\@@startlink{#1}\@@href}%
\providecommand \@@href[1]{\endgroup#1\@@endlink}%
\providecommand \@sanitize@url [0]{\catcode `\\12\catcode `\$12\catcode
  `\&12\catcode `\#12\catcode `\^12\catcode `\_12\catcode `\%12\relax}%
\providecommand \@@startlink[1]{}%
\providecommand \@@endlink[0]{}%
\providecommand \url  [0]{\begingroup\@sanitize@url \@url }%
\providecommand \@url [1]{\endgroup\@href {#1}{\urlprefix }}%
\providecommand \urlprefix  [0]{URL }%
\providecommand \Eprint [0]{\href }%
\providecommand \doibase [0]{https://doi.org/}%
\providecommand \selectlanguage [0]{\@gobble}%
\providecommand \bibinfo  [0]{\@secondoftwo}%
\providecommand \bibfield  [0]{\@secondoftwo}%
\providecommand \translation [1]{[#1]}%
\providecommand \BibitemOpen [0]{}%
\providecommand \bibitemStop [0]{}%
\providecommand \bibitemNoStop [0]{.\EOS\space}%
\providecommand \EOS [0]{\spacefactor3000\relax}%
\providecommand \BibitemShut  [1]{\csname bibitem#1\endcsname}%
\let\auto@bib@innerbib\@empty
%</preamble>
\bibitem [{\citenamefont {Gardner}\ \emph {et~al.}(2010)\citenamefont
  {Gardner}, \citenamefont {Gingras},\ and\ \citenamefont
  {Greedan}}]{RevModPhys.82.53}%
  \BibitemOpen
  \bibfield  {author} {\bibinfo {author} {\bibfnamefont {J.~S.}\ \bibnamefont
  {Gardner}}, \bibinfo {author} {\bibfnamefont {M.~J.~P.}\ \bibnamefont
  {Gingras}},\ and\ \bibinfo {author} {\bibfnamefont {J.~E.}\ \bibnamefont
  {Greedan}},\ }\bibfield  {title} {\bibinfo {title} {Magnetic pyrochlore
  oxides},\ }\href {https://doi.org/10.1103/RevModPhys.82.53} {\bibfield
  {journal} {\bibinfo  {journal} {Rev. Mod. Phys.}\ }\textbf {\bibinfo {volume}
  {82}},\ \bibinfo {pages} {53} (\bibinfo {year} {2010})}\BibitemShut {NoStop}%
\bibitem [{\citenamefont {Rau}\ and\ \citenamefont
  {Gingras}(2019)}]{annurev:10.1146}%
  \BibitemOpen
  \bibfield  {author} {\bibinfo {author} {\bibfnamefont {J.~G.}\ \bibnamefont
  {Rau}}\ and\ \bibinfo {author} {\bibfnamefont {M.~J.}\ \bibnamefont
  {Gingras}},\ }\bibfield  {title} {\bibinfo {title} {{Frustrated Quantum
  Rare-Earth Pyrochlores}},\ }\href
  {https://doi.org/https://doi.org/10.1146/annurev-conmatphys-022317-110520}
  {\bibfield  {journal} {\bibinfo  {journal} {Annual Review of Condensed Matter
  Physics}\ }\textbf {\bibinfo {volume} {10}},\ \bibinfo {pages} {357}
  (\bibinfo {year} {2019})}\BibitemShut {NoStop}%
\bibitem [{\citenamefont {Greedan}(2006)}]{GREEDAN2006444}%
  \BibitemOpen
  \bibfield  {author} {\bibinfo {author} {\bibfnamefont {J.~E.}\ \bibnamefont
  {Greedan}},\ }\bibfield  {title} {\bibinfo {title} {{Frustrated rare earth
  magnetism: Spin glasses, spin liquids and spin ices in pyrochlore oxides}},\
  }\href {https://doi.org/https://doi.org/10.1016/j.jallcom.2004.12.084}
  {\bibfield  {journal} {\bibinfo  {journal} {Journal of Alloys and Compounds}\
  }\textbf {\bibinfo {volume} {408-412}},\ \bibinfo {pages} {444} (\bibinfo
  {year} {2006})},\ \bibinfo {note} {proceedings of Rare Earths'04 in Nara,
  Japan}\BibitemShut {NoStop}%
\bibitem [{\citenamefont {Gingras}\ and\ \citenamefont
  {McClarty}(2014)}]{Gingras_2014}%
  \BibitemOpen
  \bibfield  {author} {\bibinfo {author} {\bibfnamefont {M.~J.~P.}\
  \bibnamefont {Gingras}}\ and\ \bibinfo {author} {\bibfnamefont {P.~A.}\
  \bibnamefont {McClarty}},\ }\bibfield  {title} {\bibinfo {title} {{Quantum
  spin ice: a search for gapless quantum spin liquids in pyrochlore magnets}},\
  }\href {https://doi.org/10.1088/0034-4885/77/5/056501} {\bibfield  {journal}
  {\bibinfo  {journal} {Reports on Progress in Physics}\ }\textbf {\bibinfo
  {volume} {77}},\ \bibinfo {pages} {056501} (\bibinfo {year}
  {2014})}\BibitemShut {NoStop}%
\bibitem [{\citenamefont {Ross}\ \emph {et~al.}(2011)\citenamefont {Ross},
  \citenamefont {Savary}, \citenamefont {Gaulin},\ and\ \citenamefont
  {Balents}}]{PhysRevX.1.021002}%
  \BibitemOpen
  \bibfield  {author} {\bibinfo {author} {\bibfnamefont {K.~A.}\ \bibnamefont
  {Ross}}, \bibinfo {author} {\bibfnamefont {L.}~\bibnamefont {Savary}},
  \bibinfo {author} {\bibfnamefont {B.~D.}\ \bibnamefont {Gaulin}},\ and\
  \bibinfo {author} {\bibfnamefont {L.}~\bibnamefont {Balents}},\ }\bibfield
  {title} {\bibinfo {title} {{Quantum Excitations in Quantum Spin Ice}},\
  }\href {https://doi.org/10.1103/PhysRevX.1.021002} {\bibfield  {journal}
  {\bibinfo  {journal} {Phys. Rev. X}\ }\textbf {\bibinfo {volume} {1}},\
  \bibinfo {pages} {021002} (\bibinfo {year} {2011})}\BibitemShut {NoStop}%
\bibitem [{\citenamefont {Savary}\ and\ \citenamefont
  {Balents}(2016)}]{Savary_2017}%
  \BibitemOpen
  \bibfield  {author} {\bibinfo {author} {\bibfnamefont {L.}~\bibnamefont
  {Savary}}\ and\ \bibinfo {author} {\bibfnamefont {L.}~\bibnamefont
  {Balents}},\ }\bibfield  {title} {\bibinfo {title} {Quantum spin liquids: a
  review},\ }\href {https://doi.org/10.1088/0034-4885/80/1/016502} {\bibfield
  {journal} {\bibinfo  {journal} {Reports on Progress in Physics}\ }\textbf
  {\bibinfo {volume} {80}},\ \bibinfo {pages} {016502} (\bibinfo {year}
  {2016})}\BibitemShut {NoStop}%
\bibitem [{\citenamefont {Balents}(2010)}]{balents2010}%
  \BibitemOpen
  \bibfield  {author} {\bibinfo {author} {\bibfnamefont {L.}~\bibnamefont
  {Balents}},\ }\bibfield  {title} {\bibinfo {title} {Spin liquids in
  frustrated magnets},\ }\href {https://doi.org/10.1038/nature08917} {\bibfield
   {journal} {\bibinfo  {journal} {Nature}\ }\textbf {\bibinfo {volume}
  {464}},\ \bibinfo {pages} {199} (\bibinfo {year} {2010})}\BibitemShut
  {NoStop}%
\bibitem [{\citenamefont {Castelnovo}\ \emph {et~al.}(2008)\citenamefont
  {Castelnovo}, \citenamefont {Moessner},\ and\ \citenamefont
  {Sondhi}}]{castelnovo2008}%
  \BibitemOpen
  \bibfield  {author} {\bibinfo {author} {\bibfnamefont {C.}~\bibnamefont
  {Castelnovo}}, \bibinfo {author} {\bibfnamefont {R.}~\bibnamefont
  {Moessner}},\ and\ \bibinfo {author} {\bibfnamefont {S.~L.}\ \bibnamefont
  {Sondhi}},\ }\bibfield  {title} {\bibinfo {title} {Magnetic monopoles in spin
  ice},\ }\href {https://doi.org/10.1038/nature06433} {\bibfield  {journal}
  {\bibinfo  {journal} {Nature}\ }\textbf {\bibinfo {volume} {451}},\ \bibinfo
  {pages} {42} (\bibinfo {year} {2008})}\BibitemShut {NoStop}%
\bibitem [{\citenamefont {Ramirez}\ \emph {et~al.}(1999)\citenamefont
  {Ramirez}, \citenamefont {Hayashi}, \citenamefont {Cava}, \citenamefont
  {Siddharthan},\ and\ \citenamefont {Shastry}}]{ramirez1999}%
  \BibitemOpen
  \bibfield  {author} {\bibinfo {author} {\bibfnamefont {A.~P.}\ \bibnamefont
  {Ramirez}}, \bibinfo {author} {\bibfnamefont {A.}~\bibnamefont {Hayashi}},
  \bibinfo {author} {\bibfnamefont {R.~J.}\ \bibnamefont {Cava}}, \bibinfo
  {author} {\bibfnamefont {R.}~\bibnamefont {Siddharthan}},\ and\ \bibinfo
  {author} {\bibfnamefont {B.~S.}\ \bibnamefont {Shastry}},\ }\bibfield
  {title} {\bibinfo {title} {Zero-point entropy in `spin ice'},\ }\href
  {https://doi.org/10.1038/20619} {\bibfield  {journal} {\bibinfo  {journal}
  {Nature}\ }\textbf {\bibinfo {volume} {399}},\ \bibinfo {pages} {333}
  (\bibinfo {year} {1999})}\BibitemShut {NoStop}%
\bibitem [{\citenamefont {Huse}\ \emph {et~al.}(2003)\citenamefont {Huse},
  \citenamefont {Krauth}, \citenamefont {Moessner},\ and\ \citenamefont
  {Sondhi}}]{huse2003}%
  \BibitemOpen
  \bibfield  {author} {\bibinfo {author} {\bibfnamefont {D.~A.}\ \bibnamefont
  {Huse}}, \bibinfo {author} {\bibfnamefont {W.}~\bibnamefont {Krauth}},
  \bibinfo {author} {\bibfnamefont {R.}~\bibnamefont {Moessner}},\ and\
  \bibinfo {author} {\bibfnamefont {S.~L.}\ \bibnamefont {Sondhi}},\ }\bibfield
   {title} {\bibinfo {title} {{Coulomb and Liquid Dimer Models in Three
  Dimensions}},\ }\href {https://doi.org/10.1103/PhysRevLett.91.167004}
  {\bibfield  {journal} {\bibinfo  {journal} {Phys. Rev. Lett.}\ }\textbf
  {\bibinfo {volume} {91}},\ \bibinfo {pages} {167004} (\bibinfo {year}
  {2003})}\BibitemShut {NoStop}%
\bibitem [{\citenamefont {Benton}\ \emph {et~al.}(2012)\citenamefont {Benton},
  \citenamefont {Sikora},\ and\ \citenamefont {Shannon}}]{PhysRevB.86.075154}%
  \BibitemOpen
  \bibfield  {author} {\bibinfo {author} {\bibfnamefont {O.}~\bibnamefont
  {Benton}}, \bibinfo {author} {\bibfnamefont {O.}~\bibnamefont {Sikora}},\
  and\ \bibinfo {author} {\bibfnamefont {N.}~\bibnamefont {Shannon}},\
  }\bibfield  {title} {\bibinfo {title} {{Seeing the light: Experimental
  signatures of emergent electromagnetism in a quantum spin ice}},\ }\href
  {https://doi.org/10.1103/PhysRevB.86.075154} {\bibfield  {journal} {\bibinfo
  {journal} {Phys. Rev. B}\ }\textbf {\bibinfo {volume} {86}},\ \bibinfo
  {pages} {075154} (\bibinfo {year} {2012})}\BibitemShut {NoStop}%
\bibitem [{\citenamefont {Hermele}\ \emph {et~al.}(2004)\citenamefont
  {Hermele}, \citenamefont {Fisher},\ and\ \citenamefont
  {Balents}}]{PhysRevB.69.064404}%
  \BibitemOpen
  \bibfield  {author} {\bibinfo {author} {\bibfnamefont {M.}~\bibnamefont
  {Hermele}}, \bibinfo {author} {\bibfnamefont {M.~P.~A.}\ \bibnamefont
  {Fisher}},\ and\ \bibinfo {author} {\bibfnamefont {L.}~\bibnamefont
  {Balents}},\ }\bibfield  {title} {\bibinfo {title} {{Pyrochlore photons: The
  {$U(1)$} spin liquid in a {$S=\frac{1}{2}$} three-dimensional frustrated
  magnet}},\ }\href {https://doi.org/10.1103/PhysRevB.69.064404} {\bibfield
  {journal} {\bibinfo  {journal} {Phys. Rev. B}\ }\textbf {\bibinfo {volume}
  {69}},\ \bibinfo {pages} {064404} (\bibinfo {year} {2004})}\BibitemShut
  {NoStop}%
\bibitem [{\citenamefont {Savary}\ and\ \citenamefont
  {Balents}(2012)}]{PhysRevLett.108.037202}%
  \BibitemOpen
  \bibfield  {author} {\bibinfo {author} {\bibfnamefont {L.}~\bibnamefont
  {Savary}}\ and\ \bibinfo {author} {\bibfnamefont {L.}~\bibnamefont
  {Balents}},\ }\bibfield  {title} {\bibinfo {title} {{Coulombic Quantum
  Liquids in Spin-{$1/2$} Pyrochlores}},\ }\href
  {https://doi.org/10.1103/PhysRevLett.108.037202} {\bibfield  {journal}
  {\bibinfo  {journal} {Phys. Rev. Lett.}\ }\textbf {\bibinfo {volume} {108}},\
  \bibinfo {pages} {037202} (\bibinfo {year} {2012})}\BibitemShut {NoStop}%
\bibitem [{\citenamefont {Smith}\ \emph {et~al.}(2022)\citenamefont {Smith},
  \citenamefont {Benton}, \citenamefont {Yahne}, \citenamefont {Placke},
  \citenamefont {Sch\"afer}, \citenamefont {Gaudet}, \citenamefont {Dudemaine},
  \citenamefont {Fitterman}, \citenamefont {Beare}, \citenamefont {Wildes},
  \citenamefont {Bhattacharya}, \citenamefont {DeLazzer}, \citenamefont
  {Buhariwalla}, \citenamefont {Butch}, \citenamefont {Movshovich},
  \citenamefont {Garrett}, \citenamefont {Marjerrison}, \citenamefont {Clancy},
  \citenamefont {Kermarrec}, \citenamefont {Luke}, \citenamefont {Bianchi},
  \citenamefont {Ross},\ and\ \citenamefont {Gaulin}}]{PhysRevX.12.021015}%
  \BibitemOpen
  \bibfield  {author} {\bibinfo {author} {\bibfnamefont {E.~M.}\ \bibnamefont
  {Smith}}, \bibinfo {author} {\bibfnamefont {O.}~\bibnamefont {Benton}},
  \bibinfo {author} {\bibfnamefont {D.~R.}\ \bibnamefont {Yahne}}, \bibinfo
  {author} {\bibfnamefont {B.}~\bibnamefont {Placke}}, \bibinfo {author}
  {\bibfnamefont {R.}~\bibnamefont {Sch\"afer}}, \bibinfo {author}
  {\bibfnamefont {J.}~\bibnamefont {Gaudet}}, \bibinfo {author} {\bibfnamefont
  {J.}~\bibnamefont {Dudemaine}}, \bibinfo {author} {\bibfnamefont
  {A.}~\bibnamefont {Fitterman}}, \bibinfo {author} {\bibfnamefont
  {J.}~\bibnamefont {Beare}}, \bibinfo {author} {\bibfnamefont {A.~R.}\
  \bibnamefont {Wildes}}, \bibinfo {author} {\bibfnamefont {S.}~\bibnamefont
  {Bhattacharya}}, \bibinfo {author} {\bibfnamefont {T.}~\bibnamefont
  {DeLazzer}}, \bibinfo {author} {\bibfnamefont {C.~R.~C.}\ \bibnamefont
  {Buhariwalla}}, \bibinfo {author} {\bibfnamefont {N.~P.}\ \bibnamefont
  {Butch}}, \bibinfo {author} {\bibfnamefont {R.}~\bibnamefont {Movshovich}},
  \bibinfo {author} {\bibfnamefont {J.~D.}\ \bibnamefont {Garrett}}, \bibinfo
  {author} {\bibfnamefont {C.~A.}\ \bibnamefont {Marjerrison}}, \bibinfo
  {author} {\bibfnamefont {J.~P.}\ \bibnamefont {Clancy}}, \bibinfo {author}
  {\bibfnamefont {E.}~\bibnamefont {Kermarrec}}, \bibinfo {author}
  {\bibfnamefont {G.~M.}\ \bibnamefont {Luke}}, \bibinfo {author}
  {\bibfnamefont {A.~D.}\ \bibnamefont {Bianchi}}, \bibinfo {author}
  {\bibfnamefont {K.~A.}\ \bibnamefont {Ross}},\ and\ \bibinfo {author}
  {\bibfnamefont {B.~D.}\ \bibnamefont {Gaulin}},\ }\bibfield  {title}
  {\bibinfo {title} {{Case for a {${\mathrm{U}(1)}_{\ensuremath{\pi}}$} Quantum
  Spin Liquid Ground State in the Dipole-Octupole Pyrochlore
  {${\mathrm{Ce}}_{2}{\mathrm{Zr}}_{2}{\mathrm{O}}_{7}$}}},\ }\href
  {https://doi.org/10.1103/PhysRevX.12.021015} {\bibfield  {journal} {\bibinfo
  {journal} {Phys. Rev. X}\ }\textbf {\bibinfo {volume} {12}},\ \bibinfo
  {pages} {021015} (\bibinfo {year} {2022})}\BibitemShut {NoStop}%
\bibitem [{\citenamefont {Bhardwaj}\ \emph {et~al.}(2022)\citenamefont
  {Bhardwaj}, \citenamefont {Zhang}, \citenamefont {Yan}, \citenamefont
  {Moessner}, \citenamefont {Nevidomskyy},\ and\ \citenamefont
  {Changlani}}]{bhardwaj2022}%
  \BibitemOpen
  \bibfield  {author} {\bibinfo {author} {\bibfnamefont {A.}~\bibnamefont
  {Bhardwaj}}, \bibinfo {author} {\bibfnamefont {S.}~\bibnamefont {Zhang}},
  \bibinfo {author} {\bibfnamefont {H.}~\bibnamefont {Yan}}, \bibinfo {author}
  {\bibfnamefont {R.}~\bibnamefont {Moessner}}, \bibinfo {author}
  {\bibfnamefont {A.~H.}\ \bibnamefont {Nevidomskyy}},\ and\ \bibinfo {author}
  {\bibfnamefont {H.~J.}\ \bibnamefont {Changlani}},\ }\bibfield  {title}
  {\bibinfo {title} {{Sleuthing out exotic quantum spin liquidity in the
  pyrochlore magnet Ce2Zr2O7}},\ }\href
  {https://doi.org/10.1038/s41535-022-00458-2} {\bibfield  {journal} {\bibinfo
  {journal} {npj Quantum Materials}\ }\textbf {\bibinfo {volume} {7}},\
  \bibinfo {pages} {51} (\bibinfo {year} {2022})}\BibitemShut {NoStop}%
\bibitem [{\citenamefont {Yahne}\ \emph {et~al.}(2024)\citenamefont {Yahne},
  \citenamefont {Placke}, \citenamefont {Sch\"afer}, \citenamefont {Benton},
  \citenamefont {Moessner}, \citenamefont {Powell}, \citenamefont {Kolis},
  \citenamefont {Pasco}, \citenamefont {May}, \citenamefont {Frontzek},
  \citenamefont {Smith}, \citenamefont {Gaulin}, \citenamefont {Calder},\ and\
  \citenamefont {Ross}}]{PhysRevX.14.011005}%
  \BibitemOpen
  \bibfield  {author} {\bibinfo {author} {\bibfnamefont {D.~R.}\ \bibnamefont
  {Yahne}}, \bibinfo {author} {\bibfnamefont {B.}~\bibnamefont {Placke}},
  \bibinfo {author} {\bibfnamefont {R.}~\bibnamefont {Sch\"afer}}, \bibinfo
  {author} {\bibfnamefont {O.}~\bibnamefont {Benton}}, \bibinfo {author}
  {\bibfnamefont {R.}~\bibnamefont {Moessner}}, \bibinfo {author}
  {\bibfnamefont {M.}~\bibnamefont {Powell}}, \bibinfo {author} {\bibfnamefont
  {J.~W.}\ \bibnamefont {Kolis}}, \bibinfo {author} {\bibfnamefont {C.~M.}\
  \bibnamefont {Pasco}}, \bibinfo {author} {\bibfnamefont {A.~F.}\ \bibnamefont
  {May}}, \bibinfo {author} {\bibfnamefont {M.~D.}\ \bibnamefont {Frontzek}},
  \bibinfo {author} {\bibfnamefont {E.~M.}\ \bibnamefont {Smith}}, \bibinfo
  {author} {\bibfnamefont {B.~D.}\ \bibnamefont {Gaulin}}, \bibinfo {author}
  {\bibfnamefont {S.}~\bibnamefont {Calder}},\ and\ \bibinfo {author}
  {\bibfnamefont {K.~A.}\ \bibnamefont {Ross}},\ }\bibfield  {title} {\bibinfo
  {title} {{Dipolar Spin Ice Regime Proximate to an All-In-All-Out N\'eel
  Ground State in the Dipolar-Octupolar Pyrochlore
  {${\mathrm{Ce}}_{2}{\mathrm{Sn}}_{2}{\mathrm{O}}_{7}$}}},\ }\href
  {https://doi.org/10.1103/PhysRevX.14.011005} {\bibfield  {journal} {\bibinfo
  {journal} {Phys. Rev. X}\ }\textbf {\bibinfo {volume} {14}},\ \bibinfo
  {pages} {011005} (\bibinfo {year} {2024})}\BibitemShut {NoStop}%
\bibitem [{\citenamefont {Sibille}\ \emph {et~al.}(2020)\citenamefont
  {Sibille}, \citenamefont {Gauthier}, \citenamefont {Lhotel}, \citenamefont
  {Porée}, \citenamefont {Pomjakushin}, \citenamefont {Ewings}, \citenamefont
  {Perring}, \citenamefont {Ollivier}, \citenamefont {Wildes}, \citenamefont
  {Ritter}, \citenamefont {Hansen}, \citenamefont {Keen}, \citenamefont
  {Nilsen}, \citenamefont {Keller}, \citenamefont {Petit},\ and\ \citenamefont
  {Fennell}}]{sibille_quantum_2020}%
  \BibitemOpen
  \bibfield  {author} {\bibinfo {author} {\bibfnamefont {R.}~\bibnamefont
  {Sibille}}, \bibinfo {author} {\bibfnamefont {N.}~\bibnamefont {Gauthier}},
  \bibinfo {author} {\bibfnamefont {E.}~\bibnamefont {Lhotel}}, \bibinfo
  {author} {\bibfnamefont {V.}~\bibnamefont {Porée}}, \bibinfo {author}
  {\bibfnamefont {V.}~\bibnamefont {Pomjakushin}}, \bibinfo {author}
  {\bibfnamefont {R.~A.}\ \bibnamefont {Ewings}}, \bibinfo {author}
  {\bibfnamefont {T.~G.}\ \bibnamefont {Perring}}, \bibinfo {author}
  {\bibfnamefont {J.}~\bibnamefont {Ollivier}}, \bibinfo {author}
  {\bibfnamefont {A.}~\bibnamefont {Wildes}}, \bibinfo {author} {\bibfnamefont
  {C.}~\bibnamefont {Ritter}}, \bibinfo {author} {\bibfnamefont {T.~C.}\
  \bibnamefont {Hansen}}, \bibinfo {author} {\bibfnamefont {D.~A.}\
  \bibnamefont {Keen}}, \bibinfo {author} {\bibfnamefont {G.~J.}\ \bibnamefont
  {Nilsen}}, \bibinfo {author} {\bibfnamefont {L.}~\bibnamefont {Keller}},
  \bibinfo {author} {\bibfnamefont {S.}~\bibnamefont {Petit}},\ and\ \bibinfo
  {author} {\bibfnamefont {T.}~\bibnamefont {Fennell}},\ }\bibfield  {title}
  {\bibinfo {title} {A quantum liquid of magnetic octupoles on the pyrochlore
  lattice},\ }\href {https://doi.org/10.1038/s41567-020-0827-7} {\bibfield
  {journal} {\bibinfo  {journal} {Nature Physics}\ }\textbf {\bibinfo {volume}
  {16}},\ \bibinfo {pages} {546} (\bibinfo {year} {2020})}\BibitemShut
  {NoStop}%
\bibitem [{\citenamefont {Gao}\ \emph {et~al.}(2019)\citenamefont {Gao},
  \citenamefont {Chen}, \citenamefont {Tam}, \citenamefont {Huang},
  \citenamefont {Sasmal}, \citenamefont {Adroja}, \citenamefont {Ye},
  \citenamefont {Cao}, \citenamefont {Sala}, \citenamefont {Stone},
  \citenamefont {Baines}, \citenamefont {Verezhak}, \citenamefont {Hu},
  \citenamefont {Chung}, \citenamefont {Xu}, \citenamefont {Cheong},
  \citenamefont {Nallaiyan}, \citenamefont {Spagna}, \citenamefont {Maple},
  \citenamefont {Nevidomskyy}, \citenamefont {Morosan}, \citenamefont {Chen},\
  and\ \citenamefont {Dai}}]{gao_experimental_2019}%
  \BibitemOpen
  \bibfield  {author} {\bibinfo {author} {\bibfnamefont {B.}~\bibnamefont
  {Gao}}, \bibinfo {author} {\bibfnamefont {T.}~\bibnamefont {Chen}}, \bibinfo
  {author} {\bibfnamefont {D.~W.}\ \bibnamefont {Tam}}, \bibinfo {author}
  {\bibfnamefont {C.-L.}\ \bibnamefont {Huang}}, \bibinfo {author}
  {\bibfnamefont {K.}~\bibnamefont {Sasmal}}, \bibinfo {author} {\bibfnamefont
  {D.~T.}\ \bibnamefont {Adroja}}, \bibinfo {author} {\bibfnamefont
  {F.}~\bibnamefont {Ye}}, \bibinfo {author} {\bibfnamefont {H.}~\bibnamefont
  {Cao}}, \bibinfo {author} {\bibfnamefont {G.}~\bibnamefont {Sala}}, \bibinfo
  {author} {\bibfnamefont {M.~B.}\ \bibnamefont {Stone}}, \bibinfo {author}
  {\bibfnamefont {C.}~\bibnamefont {Baines}}, \bibinfo {author} {\bibfnamefont
  {J.~A.~T.}\ \bibnamefont {Verezhak}}, \bibinfo {author} {\bibfnamefont
  {H.}~\bibnamefont {Hu}}, \bibinfo {author} {\bibfnamefont {J.-H.}\
  \bibnamefont {Chung}}, \bibinfo {author} {\bibfnamefont {X.}~\bibnamefont
  {Xu}}, \bibinfo {author} {\bibfnamefont {S.-W.}\ \bibnamefont {Cheong}},
  \bibinfo {author} {\bibfnamefont {M.}~\bibnamefont {Nallaiyan}}, \bibinfo
  {author} {\bibfnamefont {S.}~\bibnamefont {Spagna}}, \bibinfo {author}
  {\bibfnamefont {M.~B.}\ \bibnamefont {Maple}}, \bibinfo {author}
  {\bibfnamefont {A.~H.}\ \bibnamefont {Nevidomskyy}}, \bibinfo {author}
  {\bibfnamefont {E.}~\bibnamefont {Morosan}}, \bibinfo {author} {\bibfnamefont
  {G.}~\bibnamefont {Chen}},\ and\ \bibinfo {author} {\bibfnamefont
  {P.}~\bibnamefont {Dai}},\ }\bibfield  {title} {\bibinfo {title}
  {{Experimental signatures of a three-dimensional quantum spin liquid in
  effective spin-1/2 {Ce2Zr2O7} pyrochlore}},\ }\href
  {https://doi.org/10.1038/s41567-019-0577-6} {\bibfield  {journal} {\bibinfo
  {journal} {Nature Physics}\ }\textbf {\bibinfo {volume} {15}},\ \bibinfo
  {pages} {1052} (\bibinfo {year} {2019})}\BibitemShut {NoStop}%
\bibitem [{\citenamefont {Gao}\ \emph {et~al.}(2024)\citenamefont {Gao},
  \citenamefont {Desrochers}, \citenamefont {Tam}, \citenamefont {Steffens},
  \citenamefont {Hiess}, \citenamefont {Su}, \citenamefont {Cheong},
  \citenamefont {Kim},\ and\ \citenamefont {Dai}}]{gao2024}%
  \BibitemOpen
  \bibfield  {author} {\bibinfo {author} {\bibfnamefont {B.}~\bibnamefont
  {Gao}}, \bibinfo {author} {\bibfnamefont {F.}~\bibnamefont {Desrochers}},
  \bibinfo {author} {\bibfnamefont {D.~W.}\ \bibnamefont {Tam}}, \bibinfo
  {author} {\bibfnamefont {P.}~\bibnamefont {Steffens}}, \bibinfo {author}
  {\bibfnamefont {A.}~\bibnamefont {Hiess}}, \bibinfo {author} {\bibfnamefont
  {Y.}~\bibnamefont {Su}}, \bibinfo {author} {\bibfnamefont {S.-W.}\
  \bibnamefont {Cheong}}, \bibinfo {author} {\bibfnamefont {Y.~B.}\
  \bibnamefont {Kim}},\ and\ \bibinfo {author} {\bibfnamefont {P.}~\bibnamefont
  {Dai}},\ }\href@noop {} {\bibinfo {title} {Emergent photons and
  fractionalized excitations in a quantum spin liquid}} (\bibinfo {year}
  {2024}),\ \Eprint {https://arxiv.org/abs/2404.04207} {arXiv:2404.04207
  [cond-mat.str-el]} \BibitemShut {NoStop}%
\bibitem [{\citenamefont {Clark}\ \emph {et~al.}(2014)\citenamefont {Clark},
  \citenamefont {Nilsen}, \citenamefont {Kermarrec}, \citenamefont {Ehlers},
  \citenamefont {Knight}, \citenamefont {Harrison}, \citenamefont {Attfield},\
  and\ \citenamefont {Gaulin}}]{clark2014}%
  \BibitemOpen
  \bibfield  {author} {\bibinfo {author} {\bibfnamefont {L.}~\bibnamefont
  {Clark}}, \bibinfo {author} {\bibfnamefont {G.~J.}\ \bibnamefont {Nilsen}},
  \bibinfo {author} {\bibfnamefont {E.}~\bibnamefont {Kermarrec}}, \bibinfo
  {author} {\bibfnamefont {G.}~\bibnamefont {Ehlers}}, \bibinfo {author}
  {\bibfnamefont {K.~S.}\ \bibnamefont {Knight}}, \bibinfo {author}
  {\bibfnamefont {A.}~\bibnamefont {Harrison}}, \bibinfo {author}
  {\bibfnamefont {J.~P.}\ \bibnamefont {Attfield}},\ and\ \bibinfo {author}
  {\bibfnamefont {B.~D.}\ \bibnamefont {Gaulin}},\ }\bibfield  {title}
  {\bibinfo {title} {{From Spin Glass to Quantum Spin Liquid Ground States in
  Molybdate Pyrochlores}},\ }\href
  {https://doi.org/10.1103/PhysRevLett.113.117201} {\bibfield  {journal}
  {\bibinfo  {journal} {Phys. Rev. Lett.}\ }\textbf {\bibinfo {volume} {113}},\
  \bibinfo {pages} {117201} (\bibinfo {year} {2014})}\BibitemShut {NoStop}%
\bibitem [{\citenamefont {Iqbal}\ \emph {et~al.}(2017)\citenamefont {Iqbal},
  \citenamefont {M\"uller}, \citenamefont {Riedl}, \citenamefont {Reuther},
  \citenamefont {Rachel}, \citenamefont {Valent\'{\i}}, \citenamefont
  {Gingras}, \citenamefont {Thomale},\ and\ \citenamefont
  {Jeschke}}]{PhysRevMaterials.1.071201}%
  \BibitemOpen
  \bibfield  {author} {\bibinfo {author} {\bibfnamefont {Y.}~\bibnamefont
  {Iqbal}}, \bibinfo {author} {\bibfnamefont {T.}~\bibnamefont {M\"uller}},
  \bibinfo {author} {\bibfnamefont {K.}~\bibnamefont {Riedl}}, \bibinfo
  {author} {\bibfnamefont {J.}~\bibnamefont {Reuther}}, \bibinfo {author}
  {\bibfnamefont {S.}~\bibnamefont {Rachel}}, \bibinfo {author} {\bibfnamefont
  {R.}~\bibnamefont {Valent\'{\i}}}, \bibinfo {author} {\bibfnamefont
  {M.~J.~P.}\ \bibnamefont {Gingras}}, \bibinfo {author} {\bibfnamefont
  {R.}~\bibnamefont {Thomale}},\ and\ \bibinfo {author} {\bibfnamefont {H.~O.}\
  \bibnamefont {Jeschke}},\ }\bibfield  {title} {\bibinfo {title} {{Signatures
  of a gearwheel quantum spin liquid in a spin-{$\frac{1}{2}$} pyrochlore
  molybdate Heisenberg antiferromagnet}},\ }\href
  {https://doi.org/10.1103/PhysRevMaterials.1.071201} {\bibfield  {journal}
  {\bibinfo  {journal} {Phys. Rev. Mater.}\ }\textbf {\bibinfo {volume} {1}},\
  \bibinfo {pages} {071201} (\bibinfo {year} {2017})}\BibitemShut {NoStop}%
\bibitem [{\citenamefont {Canals}\ and\ \citenamefont
  {Lacroix}(1998)}]{PhysRevLett.80.2933}%
  \BibitemOpen
  \bibfield  {author} {\bibinfo {author} {\bibfnamefont {B.}~\bibnamefont
  {Canals}}\ and\ \bibinfo {author} {\bibfnamefont {C.}~\bibnamefont
  {Lacroix}},\ }\bibfield  {title} {\bibinfo {title} {{Pyrochlore
  Antiferromagnet: A Three-Dimensional Quantum Spin Liquid}},\ }\href
  {https://doi.org/10.1103/PhysRevLett.80.2933} {\bibfield  {journal} {\bibinfo
   {journal} {Phys. Rev. Lett.}\ }\textbf {\bibinfo {volume} {80}},\ \bibinfo
  {pages} {2933} (\bibinfo {year} {1998})}\BibitemShut {NoStop}%
\bibitem [{\citenamefont {Benton}\ \emph {et~al.}(2018)\citenamefont {Benton},
  \citenamefont {Jaubert}, \citenamefont {Singh}, \citenamefont {Oitmaa},\ and\
  \citenamefont {Shannon}}]{PhysRevLett.121.067201}%
  \BibitemOpen
  \bibfield  {author} {\bibinfo {author} {\bibfnamefont {O.}~\bibnamefont
  {Benton}}, \bibinfo {author} {\bibfnamefont {L.~D.~C.}\ \bibnamefont
  {Jaubert}}, \bibinfo {author} {\bibfnamefont {R.~R.~P.}\ \bibnamefont
  {Singh}}, \bibinfo {author} {\bibfnamefont {J.}~\bibnamefont {Oitmaa}},\ and\
  \bibinfo {author} {\bibfnamefont {N.}~\bibnamefont {Shannon}},\ }\bibfield
  {title} {\bibinfo {title} {{Quantum Spin Ice with Frustrated Transverse
  Exchange: From a {$\ensuremath{\pi}$}-Flux Phase to a Nematic Quantum Spin
  Liquid}},\ }\href {https://doi.org/10.1103/PhysRevLett.121.067201} {\bibfield
   {journal} {\bibinfo  {journal} {Phys. Rev. Lett.}\ }\textbf {\bibinfo
  {volume} {121}},\ \bibinfo {pages} {067201} (\bibinfo {year}
  {2018})}\BibitemShut {NoStop}%
\bibitem [{\citenamefont {Taillefumier}\ \emph {et~al.}(2017)\citenamefont
  {Taillefumier}, \citenamefont {Benton}, \citenamefont {Yan}, \citenamefont
  {Jaubert},\ and\ \citenamefont {Shannon}}]{PhysRevX.7.041057}%
  \BibitemOpen
  \bibfield  {author} {\bibinfo {author} {\bibfnamefont {M.}~\bibnamefont
  {Taillefumier}}, \bibinfo {author} {\bibfnamefont {O.}~\bibnamefont
  {Benton}}, \bibinfo {author} {\bibfnamefont {H.}~\bibnamefont {Yan}},
  \bibinfo {author} {\bibfnamefont {L.~D.~C.}\ \bibnamefont {Jaubert}},\ and\
  \bibinfo {author} {\bibfnamefont {N.}~\bibnamefont {Shannon}},\ }\bibfield
  {title} {\bibinfo {title} {{Competing Spin Liquids and Hidden Spin-Nematic
  Order in Spin Ice with Frustrated Transverse Exchange}},\ }\href
  {https://doi.org/10.1103/PhysRevX.7.041057} {\bibfield  {journal} {\bibinfo
  {journal} {Phys. Rev. X}\ }\textbf {\bibinfo {volume} {7}},\ \bibinfo {pages}
  {041057} (\bibinfo {year} {2017})}\BibitemShut {NoStop}%
\bibitem [{\citenamefont {Huang}\ \emph {et~al.}(2016)\citenamefont {Huang},
  \citenamefont {Chen}, \citenamefont {Deng}, \citenamefont {Prokof'ev},\ and\
  \citenamefont {Svistunov}}]{PhysRevLett.116.177203}%
  \BibitemOpen
  \bibfield  {author} {\bibinfo {author} {\bibfnamefont {Y.}~\bibnamefont
  {Huang}}, \bibinfo {author} {\bibfnamefont {K.}~\bibnamefont {Chen}},
  \bibinfo {author} {\bibfnamefont {Y.}~\bibnamefont {Deng}}, \bibinfo {author}
  {\bibfnamefont {N.}~\bibnamefont {Prokof'ev}},\ and\ \bibinfo {author}
  {\bibfnamefont {B.}~\bibnamefont {Svistunov}},\ }\bibfield  {title} {\bibinfo
  {title} {{Spin-Ice State of the Quantum Heisenberg Antiferromagnet on the
  Pyrochlore Lattice}},\ }\href
  {https://doi.org/10.1103/PhysRevLett.116.177203} {\bibfield  {journal}
  {\bibinfo  {journal} {Phys. Rev. Lett.}\ }\textbf {\bibinfo {volume} {116}},\
  \bibinfo {pages} {177203} (\bibinfo {year} {2016})}\BibitemShut {NoStop}%
\bibitem [{\citenamefont {Hagym\'asi}\ \emph {et~al.}(2021)\citenamefont
  {Hagym\'asi}, \citenamefont {Sch\"afer}, \citenamefont {Moessner},\ and\
  \citenamefont {Luitz}}]{PhysRevLett.126.117204}%
  \BibitemOpen
  \bibfield  {author} {\bibinfo {author} {\bibfnamefont {I.}~\bibnamefont
  {Hagym\'asi}}, \bibinfo {author} {\bibfnamefont {R.}~\bibnamefont
  {Sch\"afer}}, \bibinfo {author} {\bibfnamefont {R.}~\bibnamefont
  {Moessner}},\ and\ \bibinfo {author} {\bibfnamefont {D.~J.}\ \bibnamefont
  {Luitz}},\ }\bibfield  {title} {\bibinfo {title} {{Possible Inversion
  Symmetry Breaking in the {$S=1/2$} Pyrochlore Heisenberg Magnet}},\ }\href
  {https://doi.org/10.1103/PhysRevLett.126.117204} {\bibfield  {journal}
  {\bibinfo  {journal} {Phys. Rev. Lett.}\ }\textbf {\bibinfo {volume} {126}},\
  \bibinfo {pages} {117204} (\bibinfo {year} {2021})}\BibitemShut {NoStop}%
\bibitem [{\citenamefont {Tsunetsugu}(2001)}]{PhysRevB.65.024415}%
  \BibitemOpen
  \bibfield  {author} {\bibinfo {author} {\bibfnamefont {H.}~\bibnamefont
  {Tsunetsugu}},\ }\bibfield  {title} {\bibinfo {title} {{Spin-singlet order in
  a pyrochlore antiferromagnet}},\ }\href
  {https://doi.org/10.1103/PhysRevB.65.024415} {\bibfield  {journal} {\bibinfo
  {journal} {Phys. Rev. B}\ }\textbf {\bibinfo {volume} {65}},\ \bibinfo
  {pages} {024415} (\bibinfo {year} {2001})}\BibitemShut {NoStop}%
\bibitem [{\citenamefont {Isoda}\ and\ \citenamefont
  {Mori}(1998)}]{doi:10.1143/JPSJ.67.4022}%
  \BibitemOpen
  \bibfield  {author} {\bibinfo {author} {\bibfnamefont {M.}~\bibnamefont
  {Isoda}}\ and\ \bibinfo {author} {\bibfnamefont {S.}~\bibnamefont {Mori}},\
  }\bibfield  {title} {\bibinfo {title} {{Valence-Bond Crystal and Anisotropic
  Excitation Spectrum on 3-Dimensionally Frustrated Pyrochlore}},\ }\href
  {https://doi.org/10.1143/JPSJ.67.4022} {\bibfield  {journal} {\bibinfo
  {journal} {Journal of the Physical Society of Japan}\ }\textbf {\bibinfo
  {volume} {67}},\ \bibinfo {pages} {4022} (\bibinfo {year}
  {1998})}\BibitemShut {NoStop}%
\bibitem [{\citenamefont {Iqbal}\ \emph {et~al.}(2019)\citenamefont {Iqbal},
  \citenamefont {M\"uller}, \citenamefont {Ghosh}, \citenamefont {Gingras},
  \citenamefont {Jeschke}, \citenamefont {Rachel}, \citenamefont {Reuther},\
  and\ \citenamefont {Thomale}}]{PhysRevX.9.011005}%
  \BibitemOpen
  \bibfield  {author} {\bibinfo {author} {\bibfnamefont {Y.}~\bibnamefont
  {Iqbal}}, \bibinfo {author} {\bibfnamefont {T.}~\bibnamefont {M\"uller}},
  \bibinfo {author} {\bibfnamefont {P.}~\bibnamefont {Ghosh}}, \bibinfo
  {author} {\bibfnamefont {M.~J.~P.}\ \bibnamefont {Gingras}}, \bibinfo
  {author} {\bibfnamefont {H.~O.}\ \bibnamefont {Jeschke}}, \bibinfo {author}
  {\bibfnamefont {S.}~\bibnamefont {Rachel}}, \bibinfo {author} {\bibfnamefont
  {J.}~\bibnamefont {Reuther}},\ and\ \bibinfo {author} {\bibfnamefont
  {R.}~\bibnamefont {Thomale}},\ }\bibfield  {title} {\bibinfo {title}
  {{Quantum and Classical Phases of the Pyrochlore Heisenberg Model with
  Competing Interactions}},\ }\href {https://doi.org/10.1103/PhysRevX.9.011005}
  {\bibfield  {journal} {\bibinfo  {journal} {Phys. Rev. X}\ }\textbf {\bibinfo
  {volume} {9}},\ \bibinfo {pages} {011005} (\bibinfo {year}
  {2019})}\BibitemShut {NoStop}%
\bibitem [{\citenamefont {Niggemann}\ \emph {et~al.}(2022)\citenamefont
  {Niggemann}, \citenamefont {Reuther},\ and\ \citenamefont
  {Sbierski}}]{10.21468/SciPostPhys.12.5.156}%
  \BibitemOpen
  \bibfield  {author} {\bibinfo {author} {\bibfnamefont {N.}~\bibnamefont
  {Niggemann}}, \bibinfo {author} {\bibfnamefont {J.}~\bibnamefont {Reuther}},\
  and\ \bibinfo {author} {\bibfnamefont {B.}~\bibnamefont {Sbierski}},\
  }\bibfield  {title} {\bibinfo {title} {{Quantitative functional
  renormalization for three-dimensional quantum Heisenberg models}},\ }\href
  {https://doi.org/10.21468/SciPostPhys.12.5.156} {\bibfield  {journal}
  {\bibinfo  {journal} {SciPost Phys.}\ }\textbf {\bibinfo {volume} {12}},\
  \bibinfo {pages} {156} (\bibinfo {year} {2022})}\BibitemShut {NoStop}%
\bibitem [{\citenamefont {Harris}\ \emph {et~al.}(1991)\citenamefont {Harris},
  \citenamefont {Berlinsky},\ and\ \citenamefont
  {Bruder}}]{harris_ordering_1991}%
  \BibitemOpen
  \bibfield  {author} {\bibinfo {author} {\bibfnamefont {A.~B.}\ \bibnamefont
  {Harris}}, \bibinfo {author} {\bibfnamefont {A.~J.}\ \bibnamefont
  {Berlinsky}},\ and\ \bibinfo {author} {\bibfnamefont {C.}~\bibnamefont
  {Bruder}},\ }\bibfield  {title} {\bibinfo {title} {{Ordering by quantum
  fluctuations in a strongly frustrated {Heisenberg} antiferromagnet}},\ }\href
  {https://doi.org/10.1063/1.348098} {\bibfield  {journal} {\bibinfo  {journal}
  {Journal of Applied Physics}\ }\textbf {\bibinfo {volume} {69}},\ \bibinfo
  {pages} {5200} (\bibinfo {year} {1991})}\BibitemShut {NoStop}%
\bibitem [{\citenamefont {Sch\"afer}\ \emph {et~al.}(2023)\citenamefont
  {Sch\"afer}, \citenamefont {Placke}, \citenamefont {Benton},\ and\
  \citenamefont {Moessner}}]{PhysRevLett.131.096702}%
  \BibitemOpen
  \bibfield  {author} {\bibinfo {author} {\bibfnamefont {R.}~\bibnamefont
  {Sch\"afer}}, \bibinfo {author} {\bibfnamefont {B.}~\bibnamefont {Placke}},
  \bibinfo {author} {\bibfnamefont {O.}~\bibnamefont {Benton}},\ and\ \bibinfo
  {author} {\bibfnamefont {R.}~\bibnamefont {Moessner}},\ }\bibfield  {title}
  {\bibinfo {title} {{Abundance of Hard-Hexagon Crystals in the Quantum
  Pyrochlore Antiferromagnet}},\ }\href
  {https://doi.org/10.1103/PhysRevLett.131.096702} {\bibfield  {journal}
  {\bibinfo  {journal} {Phys. Rev. Lett.}\ }\textbf {\bibinfo {volume} {131}},\
  \bibinfo {pages} {096702} (\bibinfo {year} {2023})}\BibitemShut {NoStop}%
\bibitem [{\citenamefont {Kato}\ and\ \citenamefont
  {Onoda}(2015)}]{PhysRevLett.115.077202}%
  \BibitemOpen
  \bibfield  {author} {\bibinfo {author} {\bibfnamefont {Y.}~\bibnamefont
  {Kato}}\ and\ \bibinfo {author} {\bibfnamefont {S.}~\bibnamefont {Onoda}},\
  }\bibfield  {title} {\bibinfo {title} {{Numerical Evidence of Quantum Melting
  of Spin Ice: Quantum-to-Classical Crossover}},\ }\href
  {https://doi.org/10.1103/PhysRevLett.115.077202} {\bibfield  {journal}
  {\bibinfo  {journal} {Phys. Rev. Lett.}\ }\textbf {\bibinfo {volume} {115}},\
  \bibinfo {pages} {077202} (\bibinfo {year} {2015})}\BibitemShut {NoStop}%
\bibitem [{\citenamefont {Banerjee}\ \emph {et~al.}(2008)\citenamefont
  {Banerjee}, \citenamefont {Isakov}, \citenamefont {Damle},\ and\
  \citenamefont {Kim}}]{PhysRevLett.100.047208}%
  \BibitemOpen
  \bibfield  {author} {\bibinfo {author} {\bibfnamefont {A.}~\bibnamefont
  {Banerjee}}, \bibinfo {author} {\bibfnamefont {S.~V.}\ \bibnamefont
  {Isakov}}, \bibinfo {author} {\bibfnamefont {K.}~\bibnamefont {Damle}},\ and\
  \bibinfo {author} {\bibfnamefont {Y.~B.}\ \bibnamefont {Kim}},\ }\bibfield
  {title} {\bibinfo {title} {{Unusual Liquid State of Hard-Core Bosons on the
  Pyrochlore Lattice}},\ }\href
  {https://doi.org/10.1103/PhysRevLett.100.047208} {\bibfield  {journal}
  {\bibinfo  {journal} {Phys. Rev. Lett.}\ }\textbf {\bibinfo {volume} {100}},\
  \bibinfo {pages} {047208} (\bibinfo {year} {2008})}\BibitemShut {NoStop}%
\bibitem [{\citenamefont {Huang}\ \emph {et~al.}(2020)\citenamefont {Huang},
  \citenamefont {Liu}, \citenamefont {Meng}, \citenamefont {Yu}, \citenamefont
  {Deng},\ and\ \citenamefont {Chen}}]{PhysRevResearch.2.042022}%
  \BibitemOpen
  \bibfield  {author} {\bibinfo {author} {\bibfnamefont {C.-J.}\ \bibnamefont
  {Huang}}, \bibinfo {author} {\bibfnamefont {C.}~\bibnamefont {Liu}}, \bibinfo
  {author} {\bibfnamefont {Z.}~\bibnamefont {Meng}}, \bibinfo {author}
  {\bibfnamefont {Y.}~\bibnamefont {Yu}}, \bibinfo {author} {\bibfnamefont
  {Y.}~\bibnamefont {Deng}},\ and\ \bibinfo {author} {\bibfnamefont
  {G.}~\bibnamefont {Chen}},\ }\bibfield  {title} {\bibinfo {title} {{Extended
  Coulomb liquid of paired hardcore boson model on a pyrochlore lattice}},\
  }\href {https://doi.org/10.1103/PhysRevResearch.2.042022} {\bibfield
  {journal} {\bibinfo  {journal} {Phys. Rev. Res.}\ }\textbf {\bibinfo {volume}
  {2}},\ \bibinfo {pages} {042022} (\bibinfo {year} {2020})}\BibitemShut
  {NoStop}%
\bibitem [{\citenamefont {Chern}\ \emph {et~al.}(2024)\citenamefont {Chern},
  \citenamefont {Desrochers}, \citenamefont {Kim},\ and\ \citenamefont
  {Castelnovo}}]{PhysRevB.109.184421}%
  \BibitemOpen
  \bibfield  {author} {\bibinfo {author} {\bibfnamefont {L.~E.}\ \bibnamefont
  {Chern}}, \bibinfo {author} {\bibfnamefont {F.}~\bibnamefont {Desrochers}},
  \bibinfo {author} {\bibfnamefont {Y.~B.}\ \bibnamefont {Kim}},\ and\ \bibinfo
  {author} {\bibfnamefont {C.}~\bibnamefont {Castelnovo}},\ }\bibfield  {title}
  {\bibinfo {title} {{Pseudofermion functional renormalization group study of
  dipolar-octupolar pyrochlore magnets}},\ }\href
  {https://doi.org/10.1103/PhysRevB.109.184421} {\bibfield  {journal} {\bibinfo
   {journal} {Phys. Rev. B}\ }\textbf {\bibinfo {volume} {109}},\ \bibinfo
  {pages} {184421} (\bibinfo {year} {2024})}\BibitemShut {NoStop}%
\bibitem [{\citenamefont {Niggemann}\ \emph {et~al.}(2021)\citenamefont
  {Niggemann}, \citenamefont {Sbierski},\ and\ \citenamefont
  {Reuther}}]{PhysRevB.103.104431}%
  \BibitemOpen
  \bibfield  {author} {\bibinfo {author} {\bibfnamefont {N.}~\bibnamefont
  {Niggemann}}, \bibinfo {author} {\bibfnamefont {B.}~\bibnamefont
  {Sbierski}},\ and\ \bibinfo {author} {\bibfnamefont {J.}~\bibnamefont
  {Reuther}},\ }\bibfield  {title} {\bibinfo {title} {{Frustrated quantum spins
  at finite temperature: Pseudo-Majorana functional renormalization group
  approach}},\ }\href {https://doi.org/10.1103/PhysRevB.103.104431} {\bibfield
  {journal} {\bibinfo  {journal} {Phys. Rev. B}\ }\textbf {\bibinfo {volume}
  {103}},\ \bibinfo {pages} {104431} (\bibinfo {year} {2021})}\BibitemShut
  {NoStop}%
\bibitem [{\citenamefont {Müller}\ \emph {et~al.}(2024)\citenamefont
  {Müller}, \citenamefont {Kiese}, \citenamefont {Niggemann}, \citenamefont
  {Sbierski}, \citenamefont {Reuther}, \citenamefont {Trebst}, \citenamefont
  {Thomale},\ and\ \citenamefont {Iqbal}}]{Muller_2024}%
  \BibitemOpen
  \bibfield  {author} {\bibinfo {author} {\bibfnamefont {T.}~\bibnamefont
  {Müller}}, \bibinfo {author} {\bibfnamefont {D.}~\bibnamefont {Kiese}},
  \bibinfo {author} {\bibfnamefont {N.}~\bibnamefont {Niggemann}}, \bibinfo
  {author} {\bibfnamefont {B.}~\bibnamefont {Sbierski}}, \bibinfo {author}
  {\bibfnamefont {J.}~\bibnamefont {Reuther}}, \bibinfo {author} {\bibfnamefont
  {S.}~\bibnamefont {Trebst}}, \bibinfo {author} {\bibfnamefont
  {R.}~\bibnamefont {Thomale}},\ and\ \bibinfo {author} {\bibfnamefont
  {Y.}~\bibnamefont {Iqbal}},\ }\bibfield  {title} {\bibinfo {title}
  {{Pseudo-fermion functional renormalization group for spin models}},\ }\href
  {https://doi.org/10.1088/1361-6633/ad208c} {\bibfield  {journal} {\bibinfo
  {journal} {Reports on Progress in Physics}\ }\textbf {\bibinfo {volume}
  {87}},\ \bibinfo {pages} {036501} (\bibinfo {year} {2024})}\BibitemShut
  {NoStop}%
\bibitem [{\citenamefont {Schneider}\ \emph {et~al.}(2024)\citenamefont
  {Schneider}, \citenamefont {Reuther}, \citenamefont {Gonzalez}, \citenamefont
  {Sbierski},\ and\ \citenamefont {Niggemann}}]{PhysRevB.109.195109}%
  \BibitemOpen
  \bibfield  {author} {\bibinfo {author} {\bibfnamefont {B.}~\bibnamefont
  {Schneider}}, \bibinfo {author} {\bibfnamefont {J.}~\bibnamefont {Reuther}},
  \bibinfo {author} {\bibfnamefont {M.~G.}\ \bibnamefont {Gonzalez}}, \bibinfo
  {author} {\bibfnamefont {B.}~\bibnamefont {Sbierski}},\ and\ \bibinfo
  {author} {\bibfnamefont {N.}~\bibnamefont {Niggemann}},\ }\bibfield  {title}
  {\bibinfo {title} {{Temperature flow in pseudo-Majorana functional
  renormalization for quantum spins}},\ }\href
  {https://doi.org/10.1103/PhysRevB.109.195109} {\bibfield  {journal} {\bibinfo
   {journal} {Phys. Rev. B}\ }\textbf {\bibinfo {volume} {109}},\ \bibinfo
  {pages} {195109} (\bibinfo {year} {2024})}\BibitemShut {NoStop}%
\bibitem [{\citenamefont {Niggemann}\ \emph
  {et~al.}(2023{\natexlab{a}})\citenamefont {Niggemann}, \citenamefont
  {Iqbal},\ and\ \citenamefont {Reuther}}]{PhysRevLett.130.196601}%
  \BibitemOpen
  \bibfield  {author} {\bibinfo {author} {\bibfnamefont {N.}~\bibnamefont
  {Niggemann}}, \bibinfo {author} {\bibfnamefont {Y.}~\bibnamefont {Iqbal}},\
  and\ \bibinfo {author} {\bibfnamefont {J.}~\bibnamefont {Reuther}},\
  }\bibfield  {title} {\bibinfo {title} {{Quantum Effects on Unconventional
  Pinch Point Singularities}},\ }\href
  {https://doi.org/10.1103/PhysRevLett.130.196601} {\bibfield  {journal}
  {\bibinfo  {journal} {Phys. Rev. Lett.}\ }\textbf {\bibinfo {volume} {130}},\
  \bibinfo {pages} {196601} (\bibinfo {year} {2023}{\natexlab{a}})}\BibitemShut
  {NoStop}%
\bibitem [{\citenamefont {Astrakhantsev}\ \emph
  {et~al.}(2021{\natexlab{a}})\citenamefont {Astrakhantsev}, \citenamefont
  {Ferrari}, \citenamefont {Niggemann}, \citenamefont {M\"uller}, \citenamefont
  {Chauhan}, \citenamefont {Kshetrimayum}, \citenamefont {Ghosh}, \citenamefont
  {Regnault}, \citenamefont {Thomale}, \citenamefont {Reuther}, \citenamefont
  {Neupert},\ and\ \citenamefont {Iqbal}}]{PhysRevB.104.L220408}%
  \BibitemOpen
  \bibfield  {author} {\bibinfo {author} {\bibfnamefont {N.}~\bibnamefont
  {Astrakhantsev}}, \bibinfo {author} {\bibfnamefont {F.}~\bibnamefont
  {Ferrari}}, \bibinfo {author} {\bibfnamefont {N.}~\bibnamefont {Niggemann}},
  \bibinfo {author} {\bibfnamefont {T.}~\bibnamefont {M\"uller}}, \bibinfo
  {author} {\bibfnamefont {A.}~\bibnamefont {Chauhan}}, \bibinfo {author}
  {\bibfnamefont {A.}~\bibnamefont {Kshetrimayum}}, \bibinfo {author}
  {\bibfnamefont {P.}~\bibnamefont {Ghosh}}, \bibinfo {author} {\bibfnamefont
  {N.}~\bibnamefont {Regnault}}, \bibinfo {author} {\bibfnamefont
  {R.}~\bibnamefont {Thomale}}, \bibinfo {author} {\bibfnamefont
  {J.}~\bibnamefont {Reuther}}, \bibinfo {author} {\bibfnamefont
  {T.}~\bibnamefont {Neupert}},\ and\ \bibinfo {author} {\bibfnamefont
  {Y.}~\bibnamefont {Iqbal}},\ }\bibfield  {title} {\bibinfo {title} {{Pinwheel
  valence bond crystal ground state of the spin-$\frac{1}{2}$ Heisenberg
  antiferromagnet on the shuriken lattice}},\ }\href
  {https://doi.org/10.1103/PhysRevB.104.L220408} {\bibfield  {journal}
  {\bibinfo  {journal} {Phys. Rev. B}\ }\textbf {\bibinfo {volume} {104}},\
  \bibinfo {pages} {L220408} (\bibinfo {year}
  {2021}{\natexlab{a}})}\BibitemShut {NoStop}%
\bibitem [{\citenamefont {Niggemann}\ \emph
  {et~al.}(2023{\natexlab{b}})\citenamefont {Niggemann}, \citenamefont
  {Astrakhantsev}, \citenamefont {Ralko}, \citenamefont {Ferrari},
  \citenamefont {Maity}, \citenamefont {M\"uller}, \citenamefont {Richter},
  \citenamefont {Thomale}, \citenamefont {Neupert}, \citenamefont {Reuther},
  \citenamefont {Iqbal},\ and\ \citenamefont {Jeschke}}]{niggemann_decorated}%
  \BibitemOpen
  \bibfield  {author} {\bibinfo {author} {\bibfnamefont {N.}~\bibnamefont
  {Niggemann}}, \bibinfo {author} {\bibfnamefont {N.}~\bibnamefont
  {Astrakhantsev}}, \bibinfo {author} {\bibfnamefont {A.}~\bibnamefont
  {Ralko}}, \bibinfo {author} {\bibfnamefont {F.}~\bibnamefont {Ferrari}},
  \bibinfo {author} {\bibfnamefont {A.}~\bibnamefont {Maity}}, \bibinfo
  {author} {\bibfnamefont {T.}~\bibnamefont {M\"uller}}, \bibinfo {author}
  {\bibfnamefont {J.}~\bibnamefont {Richter}}, \bibinfo {author} {\bibfnamefont
  {R.}~\bibnamefont {Thomale}}, \bibinfo {author} {\bibfnamefont
  {T.}~\bibnamefont {Neupert}}, \bibinfo {author} {\bibfnamefont
  {J.}~\bibnamefont {Reuther}}, \bibinfo {author} {\bibfnamefont
  {Y.}~\bibnamefont {Iqbal}},\ and\ \bibinfo {author} {\bibfnamefont {H.~O.}\
  \bibnamefont {Jeschke}},\ }\bibfield  {title} {\bibinfo {title} {{Quantum
  paramagnetism in the decorated square-kagome antiferromagnet
  ${\mathrm{Na}}_{6}{\mathrm{Cu}}_{7}{\mathrm{BiO}}_{4}{({\mathrm{PO}}_{4})}_{4}{\mathrm{Cl}}_{3}$}},\
  }\href {https://doi.org/10.1103/PhysRevB.108.L241117} {\bibfield  {journal}
  {\bibinfo  {journal} {Phys. Rev. B}\ }\textbf {\bibinfo {volume} {108}},\
  \bibinfo {pages} {L241117} (\bibinfo {year}
  {2023}{\natexlab{b}})}\BibitemShut {NoStop}%
\bibitem [{\citenamefont {Hagymási}\ \emph {et~al.}(2024)\citenamefont
  {Hagymási}, \citenamefont {Niggemann},\ and\ \citenamefont
  {Reuther}}]{hagymasi2024}%
  \BibitemOpen
  \bibfield  {author} {\bibinfo {author} {\bibfnamefont {I.}~\bibnamefont
  {Hagymási}}, \bibinfo {author} {\bibfnamefont {N.}~\bibnamefont
  {Niggemann}},\ and\ \bibinfo {author} {\bibfnamefont {J.}~\bibnamefont
  {Reuther}},\ }\href@noop {} {\bibinfo {title} {Phase diagram of the
  antiferromagnetic $j_1$-$j_2$ spin-$1$ pyrochlore heisenberg model}}
  (\bibinfo {year} {2024}),\ \Eprint {https://arxiv.org/abs/2405.12745}
  {arXiv:2405.12745 [cond-mat.str-el]} \BibitemShut {NoStop}%
\bibitem [{\citenamefont {Bippus}\ \emph {et~al.}(2024)\citenamefont {Bippus},
  \citenamefont {Schneider},\ and\ \citenamefont {Sbierski}}]{bippus2024}%
  \BibitemOpen
  \bibfield  {author} {\bibinfo {author} {\bibfnamefont {F.}~\bibnamefont
  {Bippus}}, \bibinfo {author} {\bibfnamefont {B.}~\bibnamefont {Schneider}},\
  and\ \bibinfo {author} {\bibfnamefont {B.}~\bibnamefont {Sbierski}},\
  }\href@noop {} {\bibinfo {title} {{Pseudo-Majorana Functional Renormalization
  for Frustrated XXZ-Z Spin-1/2 Models}}} (\bibinfo {year} {2024}),\ \Eprint
  {https://arxiv.org/abs/2411.18198} {arXiv:2411.18198 [cond-mat.str-el]}
  \BibitemShut {NoStop}%
\bibitem [{\citenamefont {Sbierski}\ \emph {et~al.}(2024)\citenamefont
  {Sbierski}, \citenamefont {Bintz}, \citenamefont {Chatterjee}, \citenamefont
  {Schuler}, \citenamefont {Yao},\ and\ \citenamefont
  {Pollet}}]{PhysRevB.109.144411}%
  \BibitemOpen
  \bibfield  {author} {\bibinfo {author} {\bibfnamefont {B.}~\bibnamefont
  {Sbierski}}, \bibinfo {author} {\bibfnamefont {M.}~\bibnamefont {Bintz}},
  \bibinfo {author} {\bibfnamefont {S.}~\bibnamefont {Chatterjee}}, \bibinfo
  {author} {\bibfnamefont {M.}~\bibnamefont {Schuler}}, \bibinfo {author}
  {\bibfnamefont {N.~Y.}\ \bibnamefont {Yao}},\ and\ \bibinfo {author}
  {\bibfnamefont {L.}~\bibnamefont {Pollet}},\ }\bibfield  {title} {\bibinfo
  {title} {{Magnetism in the two-dimensional dipolar XY model}},\ }\href
  {https://doi.org/10.1103/PhysRevB.109.144411} {\bibfield  {journal} {\bibinfo
   {journal} {Phys. Rev. B}\ }\textbf {\bibinfo {volume} {109}},\ \bibinfo
  {pages} {144411} (\bibinfo {year} {2024})}\BibitemShut {NoStop}%
\bibitem [{\citenamefont {Noculak}\ \emph {et~al.}(2023)\citenamefont
  {Noculak}, \citenamefont {Lozano-G\'omez}, \citenamefont {Oitmaa},
  \citenamefont {Singh}, \citenamefont {Iqbal}, \citenamefont {Gingras},\ and\
  \citenamefont {Reuther}}]{PhysRevB.107.214414}%
  \BibitemOpen
  \bibfield  {author} {\bibinfo {author} {\bibfnamefont {V.}~\bibnamefont
  {Noculak}}, \bibinfo {author} {\bibfnamefont {D.}~\bibnamefont
  {Lozano-G\'omez}}, \bibinfo {author} {\bibfnamefont {J.}~\bibnamefont
  {Oitmaa}}, \bibinfo {author} {\bibfnamefont {R.~R.~P.}\ \bibnamefont
  {Singh}}, \bibinfo {author} {\bibfnamefont {Y.}~\bibnamefont {Iqbal}},
  \bibinfo {author} {\bibfnamefont {M.~J.~P.}\ \bibnamefont {Gingras}},\ and\
  \bibinfo {author} {\bibfnamefont {J.}~\bibnamefont {Reuther}},\ }\bibfield
  {title} {\bibinfo {title} {{Classical and quantum phases of the pyrochlore
  {$S=\frac{1}{2}$} magnet with Heisenberg and Dzyaloshinskii-Moriya
  interactions}},\ }\href {https://doi.org/10.1103/PhysRevB.107.214414}
  {\bibfield  {journal} {\bibinfo  {journal} {Phys. Rev. B}\ }\textbf {\bibinfo
  {volume} {107}},\ \bibinfo {pages} {214414} (\bibinfo {year}
  {2023})}\BibitemShut {NoStop}%
\bibitem [{\citenamefont {Lozano-Gómez}\ \emph
  {et~al.}(2024{\natexlab{a}})\citenamefont {Lozano-Gómez}, \citenamefont
  {Noculak}, \citenamefont {Oitmaa}, \citenamefont {Singh}, \citenamefont
  {Iqbal}, \citenamefont {Reuther},\ and\ \citenamefont {Gingras}}]{gomez2024}%
  \BibitemOpen
  \bibfield  {author} {\bibinfo {author} {\bibfnamefont {D.}~\bibnamefont
  {Lozano-Gómez}}, \bibinfo {author} {\bibfnamefont {V.}~\bibnamefont
  {Noculak}}, \bibinfo {author} {\bibfnamefont {J.}~\bibnamefont {Oitmaa}},
  \bibinfo {author} {\bibfnamefont {R.~R.~P.}\ \bibnamefont {Singh}}, \bibinfo
  {author} {\bibfnamefont {Y.}~\bibnamefont {Iqbal}}, \bibinfo {author}
  {\bibfnamefont {J.}~\bibnamefont {Reuther}},\ and\ \bibinfo {author}
  {\bibfnamefont {M.~J.~P.}\ \bibnamefont {Gingras}},\ }\bibfield  {title}
  {\bibinfo {title} {{Competing gauge fields and entropically driven spin
  liquid to spin liquid transition in non-Kramers pyrochlores}},\ }\href
  {https://doi.org/10.1073/pnas.2403487121} {\bibfield  {journal} {\bibinfo
  {journal} {Proceedings of the National Academy of Sciences}\ }\textbf
  {\bibinfo {volume} {121}},\ \bibinfo {pages} {e2403487121} (\bibinfo {year}
  {2024}{\natexlab{a}})}\BibitemShut {NoStop}%
\bibitem [{\citenamefont {Hering}\ \emph {et~al.}(2022)\citenamefont {Hering},
  \citenamefont {Noculak}, \citenamefont {Ferrari}, \citenamefont {Iqbal},\
  and\ \citenamefont {Reuther}}]{PhysRevB.105.054426}%
  \BibitemOpen
  \bibfield  {author} {\bibinfo {author} {\bibfnamefont {M.}~\bibnamefont
  {Hering}}, \bibinfo {author} {\bibfnamefont {V.}~\bibnamefont {Noculak}},
  \bibinfo {author} {\bibfnamefont {F.}~\bibnamefont {Ferrari}}, \bibinfo
  {author} {\bibfnamefont {Y.}~\bibnamefont {Iqbal}},\ and\ \bibinfo {author}
  {\bibfnamefont {J.}~\bibnamefont {Reuther}},\ }\bibfield  {title} {\bibinfo
  {title} {{Dimerization tendencies of the pyrochlore Heisenberg
  antiferromagnet: A functional renormalization group perspective}},\ }\href
  {https://doi.org/10.1103/PhysRevB.105.054426} {\bibfield  {journal} {\bibinfo
   {journal} {Phys. Rev. B}\ }\textbf {\bibinfo {volume} {105}},\ \bibinfo
  {pages} {054426} (\bibinfo {year} {2022})}\BibitemShut {NoStop}%
\bibitem [{\citenamefont {Hagym\'asi}\ \emph {et~al.}(2022)\citenamefont
  {Hagym\'asi}, \citenamefont {Noculak},\ and\ \citenamefont
  {Reuther}}]{PhysRevB.106.235137}%
  \BibitemOpen
  \bibfield  {author} {\bibinfo {author} {\bibfnamefont {I.}~\bibnamefont
  {Hagym\'asi}}, \bibinfo {author} {\bibfnamefont {V.}~\bibnamefont
  {Noculak}},\ and\ \bibinfo {author} {\bibfnamefont {J.}~\bibnamefont
  {Reuther}},\ }\bibfield  {title} {\bibinfo {title} {{Enhanced
  symmetry-breaking tendencies in the $S=1$ pyrochlore antiferromagnet}},\
  }\href {https://doi.org/10.1103/PhysRevB.106.235137} {\bibfield  {journal}
  {\bibinfo  {journal} {Phys. Rev. B}\ }\textbf {\bibinfo {volume} {106}},\
  \bibinfo {pages} {235137} (\bibinfo {year} {2022})}\BibitemShut {NoStop}%
\bibitem [{\citenamefont {Ghosh}\ \emph {et~al.}(2019)\citenamefont {Ghosh},
  \citenamefont {Iqbal}, \citenamefont {M{\"u}ller}, \citenamefont
  {Ponnaganti}, \citenamefont {Thomale}, \citenamefont {Narayanan},
  \citenamefont {Reuther}, \citenamefont {Gingras},\ and\ \citenamefont
  {Jeschke}}]{ghosh2019}%
  \BibitemOpen
  \bibfield  {author} {\bibinfo {author} {\bibfnamefont {P.}~\bibnamefont
  {Ghosh}}, \bibinfo {author} {\bibfnamefont {Y.}~\bibnamefont {Iqbal}},
  \bibinfo {author} {\bibfnamefont {T.}~\bibnamefont {M{\"u}ller}}, \bibinfo
  {author} {\bibfnamefont {R.~T.}\ \bibnamefont {Ponnaganti}}, \bibinfo
  {author} {\bibfnamefont {R.}~\bibnamefont {Thomale}}, \bibinfo {author}
  {\bibfnamefont {R.}~\bibnamefont {Narayanan}}, \bibinfo {author}
  {\bibfnamefont {J.}~\bibnamefont {Reuther}}, \bibinfo {author} {\bibfnamefont
  {M.~J.~P.}\ \bibnamefont {Gingras}},\ and\ \bibinfo {author} {\bibfnamefont
  {H.~O.}\ \bibnamefont {Jeschke}},\ }\bibfield  {title} {\bibinfo {title}
  {{Breathing chromium spinels: a showcase for a variety of pyrochlore
  Heisenberg Hamiltonians}},\ }\href
  {https://doi.org/10.1038/s41535-019-0202-z} {\bibfield  {journal} {\bibinfo
  {journal} {npj Quantum Materials}\ }\textbf {\bibinfo {volume} {4}},\
  \bibinfo {pages} {63} (\bibinfo {year} {2019})}\BibitemShut {NoStop}%
\bibitem [{\citenamefont {Schneider}\ \emph {et~al.}(2022)\citenamefont
  {Schneider}, \citenamefont {Kiese},\ and\ \citenamefont
  {Sbierski}}]{PhysRevB.106.235113}%
  \BibitemOpen
  \bibfield  {author} {\bibinfo {author} {\bibfnamefont {B.}~\bibnamefont
  {Schneider}}, \bibinfo {author} {\bibfnamefont {D.}~\bibnamefont {Kiese}},\
  and\ \bibinfo {author} {\bibfnamefont {B.}~\bibnamefont {Sbierski}},\
  }\bibfield  {title} {\bibinfo {title} {{Taming pseudofermion functional
  renormalization for quantum spins: Finite temperatures and the Popov-Fedotov
  trick}},\ }\href {https://doi.org/10.1103/PhysRevB.106.235113} {\bibfield
  {journal} {\bibinfo  {journal} {Phys. Rev. B}\ }\textbf {\bibinfo {volume}
  {106}},\ \bibinfo {pages} {235113} (\bibinfo {year} {2022})}\BibitemShut
  {NoStop}%
\bibitem [{\citenamefont {Astrakhantsev}\ \emph
  {et~al.}(2021{\natexlab{b}})\citenamefont {Astrakhantsev}, \citenamefont
  {Westerhout}, \citenamefont {Tiwari}, \citenamefont {Choo}, \citenamefont
  {Chen}, \citenamefont {Fischer}, \citenamefont {Carleo},\ and\ \citenamefont
  {Neupert}}]{PhysRevX.11.041021}%
  \BibitemOpen
  \bibfield  {author} {\bibinfo {author} {\bibfnamefont {N.}~\bibnamefont
  {Astrakhantsev}}, \bibinfo {author} {\bibfnamefont {T.}~\bibnamefont
  {Westerhout}}, \bibinfo {author} {\bibfnamefont {A.}~\bibnamefont {Tiwari}},
  \bibinfo {author} {\bibfnamefont {K.}~\bibnamefont {Choo}}, \bibinfo {author}
  {\bibfnamefont {A.}~\bibnamefont {Chen}}, \bibinfo {author} {\bibfnamefont
  {M.~H.}\ \bibnamefont {Fischer}}, \bibinfo {author} {\bibfnamefont
  {G.}~\bibnamefont {Carleo}},\ and\ \bibinfo {author} {\bibfnamefont
  {T.}~\bibnamefont {Neupert}},\ }\bibfield  {title} {\bibinfo {title}
  {{Broken-Symmetry Ground States of the Heisenberg Model on the Pyrochlore
  Lattice}},\ }\href {https://doi.org/10.1103/PhysRevX.11.041021} {\bibfield
  {journal} {\bibinfo  {journal} {Phys. Rev. X}\ }\textbf {\bibinfo {volume}
  {11}},\ \bibinfo {pages} {041021} (\bibinfo {year}
  {2021}{\natexlab{b}})}\BibitemShut {NoStop}%
\bibitem [{\citenamefont {Yan}\ \emph {et~al.}(2024)\citenamefont {Yan},
  \citenamefont {Benton}, \citenamefont {Nevidomskyy},\ and\ \citenamefont
  {Moessner}}]{PhysRevB.109.174421}%
  \BibitemOpen
  \bibfield  {author} {\bibinfo {author} {\bibfnamefont {H.}~\bibnamefont
  {Yan}}, \bibinfo {author} {\bibfnamefont {O.}~\bibnamefont {Benton}},
  \bibinfo {author} {\bibfnamefont {A.~H.}\ \bibnamefont {Nevidomskyy}},\ and\
  \bibinfo {author} {\bibfnamefont {R.}~\bibnamefont {Moessner}},\ }\bibfield
  {title} {\bibinfo {title} {{Classification of classical spin liquids:
  Detailed formalism and suite of examples}},\ }\href
  {https://doi.org/10.1103/PhysRevB.109.174421} {\bibfield  {journal} {\bibinfo
   {journal} {Phys. Rev. B}\ }\textbf {\bibinfo {volume} {109}},\ \bibinfo
  {pages} {174421} (\bibinfo {year} {2024})}\BibitemShut {NoStop}%
\bibitem [{\citenamefont {Harris}\ \emph {et~al.}(1998)\citenamefont {Harris},
  \citenamefont {Bramwell}, \citenamefont {Holdsworth},\ and\ \citenamefont
  {Champion}}]{PhysRevLett.81.4496}%
  \BibitemOpen
  \bibfield  {author} {\bibinfo {author} {\bibfnamefont {M.~J.}\ \bibnamefont
  {Harris}}, \bibinfo {author} {\bibfnamefont {S.~T.}\ \bibnamefont
  {Bramwell}}, \bibinfo {author} {\bibfnamefont {P.~C.~W.}\ \bibnamefont
  {Holdsworth}},\ and\ \bibinfo {author} {\bibfnamefont {J.~D.~M.}\
  \bibnamefont {Champion}},\ }\bibfield  {title} {\bibinfo {title} {{Liquid-Gas
  Critical Behavior in a Frustrated Pyrochlore Ferromagnet}},\ }\href
  {https://doi.org/10.1103/PhysRevLett.81.4496} {\bibfield  {journal} {\bibinfo
   {journal} {Phys. Rev. Lett.}\ }\textbf {\bibinfo {volume} {81}},\ \bibinfo
  {pages} {4496} (\bibinfo {year} {1998})}\BibitemShut {NoStop}%
\bibitem [{\citenamefont {Bramwell}\ and\ \citenamefont
  {Gingras}(2001)}]{doi:10.1126/science.1064761}%
  \BibitemOpen
  \bibfield  {author} {\bibinfo {author} {\bibfnamefont {S.~T.}\ \bibnamefont
  {Bramwell}}\ and\ \bibinfo {author} {\bibfnamefont {M.~J.~P.}\ \bibnamefont
  {Gingras}},\ }\bibfield  {title} {\bibinfo {title} {{Spin Ice State in
  Frustrated Magnetic Pyrochlore Materials}},\ }\href
  {https://doi.org/10.1126/science.1064761} {\bibfield  {journal} {\bibinfo
  {journal} {Science}\ }\textbf {\bibinfo {volume} {294}},\ \bibinfo {pages}
  {1495} (\bibinfo {year} {2001})}\BibitemShut {NoStop}%
\bibitem [{\citenamefont {Isakov}\ \emph {et~al.}(2004)\citenamefont {Isakov},
  \citenamefont {Gregor}, \citenamefont {Moessner},\ and\ \citenamefont
  {Sondhi}}]{PhysRevLett.93.167204}%
  \BibitemOpen
  \bibfield  {author} {\bibinfo {author} {\bibfnamefont {S.~V.}\ \bibnamefont
  {Isakov}}, \bibinfo {author} {\bibfnamefont {K.}~\bibnamefont {Gregor}},
  \bibinfo {author} {\bibfnamefont {R.}~\bibnamefont {Moessner}},\ and\
  \bibinfo {author} {\bibfnamefont {S.~L.}\ \bibnamefont {Sondhi}},\ }\bibfield
   {title} {\bibinfo {title} {{Dipolar Spin Correlations in Classical
  Pyrochlore Magnets}},\ }\href {https://doi.org/10.1103/PhysRevLett.93.167204}
  {\bibfield  {journal} {\bibinfo  {journal} {Phys. Rev. Lett.}\ }\textbf
  {\bibinfo {volume} {93}},\ \bibinfo {pages} {167204} (\bibinfo {year}
  {2004})}\BibitemShut {NoStop}%
\bibitem [{\citenamefont {Yan}\ \emph {et~al.}(2023)\citenamefont {Yan},
  \citenamefont {Benton}, \citenamefont {Moessner},\ and\ \citenamefont
  {Nevidomskyy}}]{yan2023}%
  \BibitemOpen
  \bibfield  {author} {\bibinfo {author} {\bibfnamefont {H.}~\bibnamefont
  {Yan}}, \bibinfo {author} {\bibfnamefont {O.}~\bibnamefont {Benton}},
  \bibinfo {author} {\bibfnamefont {R.}~\bibnamefont {Moessner}},\ and\
  \bibinfo {author} {\bibfnamefont {A.~H.}\ \bibnamefont {Nevidomskyy}},\
  }\href@noop {} {\bibinfo {title} {Classification of classical spin liquids:
  Typology and resulting landscape}} (\bibinfo {year} {2023}),\ \Eprint
  {https://arxiv.org/abs/2305.00155} {arXiv:2305.00155 [cond-mat.str-el]}
  \BibitemShut {NoStop}%
\bibitem [{\citenamefont {Bramwell}\ \emph {et~al.}(2001)\citenamefont
  {Bramwell}, \citenamefont {Harris}, \citenamefont {den Hertog}, \citenamefont
  {Gingras}, \citenamefont {Gardner}, \citenamefont {McMorrow}, \citenamefont
  {Wildes}, \citenamefont {Cornelius}, \citenamefont {Champion}, \citenamefont
  {Melko},\ and\ \citenamefont {Fennell}}]{PhysRevLett.87.047205}%
  \BibitemOpen
  \bibfield  {author} {\bibinfo {author} {\bibfnamefont {S.~T.}\ \bibnamefont
  {Bramwell}}, \bibinfo {author} {\bibfnamefont {M.~J.}\ \bibnamefont
  {Harris}}, \bibinfo {author} {\bibfnamefont {B.~C.}\ \bibnamefont {den
  Hertog}}, \bibinfo {author} {\bibfnamefont {M.~J.~P.}\ \bibnamefont
  {Gingras}}, \bibinfo {author} {\bibfnamefont {J.~S.}\ \bibnamefont
  {Gardner}}, \bibinfo {author} {\bibfnamefont {D.~F.}\ \bibnamefont
  {McMorrow}}, \bibinfo {author} {\bibfnamefont {A.~R.}\ \bibnamefont
  {Wildes}}, \bibinfo {author} {\bibfnamefont {A.~L.}\ \bibnamefont
  {Cornelius}}, \bibinfo {author} {\bibfnamefont {J.~D.~M.}\ \bibnamefont
  {Champion}}, \bibinfo {author} {\bibfnamefont {R.~G.}\ \bibnamefont
  {Melko}},\ and\ \bibinfo {author} {\bibfnamefont {T.}~\bibnamefont
  {Fennell}},\ }\bibfield  {title} {\bibinfo {title} {{Spin Correlations in
  {${\mathrm{Ho}}_{2}{\mathrm{Ti}}_{2}{O}_{7}$}: A Dipolar Spin Ice System}},\
  }\href {https://doi.org/10.1103/PhysRevLett.87.047205} {\bibfield  {journal}
  {\bibinfo  {journal} {Phys. Rev. Lett.}\ }\textbf {\bibinfo {volume} {87}},\
  \bibinfo {pages} {047205} (\bibinfo {year} {2001})}\BibitemShut {NoStop}%
\bibitem [{\citenamefont {Harris}\ \emph {et~al.}(1997)\citenamefont {Harris},
  \citenamefont {Bramwell}, \citenamefont {McMorrow}, \citenamefont {Zeiske},\
  and\ \citenamefont {Godfrey}}]{PhysRevLett.79.2554}%
  \BibitemOpen
  \bibfield  {author} {\bibinfo {author} {\bibfnamefont {M.~J.}\ \bibnamefont
  {Harris}}, \bibinfo {author} {\bibfnamefont {S.~T.}\ \bibnamefont
  {Bramwell}}, \bibinfo {author} {\bibfnamefont {D.~F.}\ \bibnamefont
  {McMorrow}}, \bibinfo {author} {\bibfnamefont {T.}~\bibnamefont {Zeiske}},\
  and\ \bibinfo {author} {\bibfnamefont {K.~W.}\ \bibnamefont {Godfrey}},\
  }\bibfield  {title} {\bibinfo {title} {{Geometrical Frustration in the
  Ferromagnetic Pyrochlore {${\mathrm{Ho}}_{2}{\mathrm{Ti}}_{2}{O}_{7}$}}},\
  }\href {https://doi.org/10.1103/PhysRevLett.79.2554} {\bibfield  {journal}
  {\bibinfo  {journal} {Phys. Rev. Lett.}\ }\textbf {\bibinfo {volume} {79}},\
  \bibinfo {pages} {2554} (\bibinfo {year} {1997})}\BibitemShut {NoStop}%
\bibitem [{\citenamefont {Lago}\ \emph {et~al.}(2007)\citenamefont {Lago},
  \citenamefont {Blundell},\ and\ \citenamefont {Baines}}]{Lago_2007}%
  \BibitemOpen
  \bibfield  {author} {\bibinfo {author} {\bibfnamefont {J.}~\bibnamefont
  {Lago}}, \bibinfo {author} {\bibfnamefont {S.~J.}\ \bibnamefont {Blundell}},\
  and\ \bibinfo {author} {\bibfnamefont {C.}~\bibnamefont {Baines}},\
  }\bibfield  {title} {\bibinfo {title} {{$\mu$SR investigation of spin
  dynamics in the spin-ice material Dy2Ti2O7}},\ }\href
  {https://doi.org/10.1088/0953-8984/19/32/326210} {\bibfield  {journal}
  {\bibinfo  {journal} {Journal of Physics: Condensed Matter}\ }\textbf
  {\bibinfo {volume} {19}},\ \bibinfo {pages} {326210} (\bibinfo {year}
  {2007})}\BibitemShut {NoStop}%
\bibitem [{\citenamefont {Desrochers}\ and\ \citenamefont
  {Kim}(2024)}]{PhysRevLett.132.066502}%
  \BibitemOpen
  \bibfield  {author} {\bibinfo {author} {\bibfnamefont {F.}~\bibnamefont
  {Desrochers}}\ and\ \bibinfo {author} {\bibfnamefont {Y.~B.}\ \bibnamefont
  {Kim}},\ }\bibfield  {title} {\bibinfo {title} {{Spectroscopic Signatures of
  Fractionalization in Octupolar Quantum Spin Ice}},\ }\href
  {https://doi.org/10.1103/PhysRevLett.132.066502} {\bibfield  {journal}
  {\bibinfo  {journal} {Phys. Rev. Lett.}\ }\textbf {\bibinfo {volume} {132}},\
  \bibinfo {pages} {066502} (\bibinfo {year} {2024})}\BibitemShut {NoStop}%
\bibitem [{\citenamefont {Sanders}\ \emph {et~al.}(2024)\citenamefont
  {Sanders}, \citenamefont {Yan}, \citenamefont {Castelnovo},\ and\
  \citenamefont {Nevidomskyy}}]{sanders2024}%
  \BibitemOpen
  \bibfield  {author} {\bibinfo {author} {\bibfnamefont {A.}~\bibnamefont
  {Sanders}}, \bibinfo {author} {\bibfnamefont {H.}~\bibnamefont {Yan}},
  \bibinfo {author} {\bibfnamefont {C.}~\bibnamefont {Castelnovo}},\ and\
  \bibinfo {author} {\bibfnamefont {A.~H.}\ \bibnamefont {Nevidomskyy}},\
  }\href@noop {} {\bibinfo {title} {{Experimentally tunable QED in
  dipolar-octupolar quantum spin ice}}} (\bibinfo {year} {2024}),\ \Eprint
  {https://arxiv.org/abs/2312.11641} {arXiv:2312.11641 [cond-mat.str-el]}
  \BibitemShut {NoStop}%
\bibitem [{\citenamefont {Takatsu}\ \emph {et~al.}(2016)\citenamefont
  {Takatsu}, \citenamefont {Onoda}, \citenamefont {Kittaka}, \citenamefont
  {Kasahara}, \citenamefont {Kono}, \citenamefont {Sakakibara}, \citenamefont
  {Kato}, \citenamefont {F\aa{}k}, \citenamefont {Ollivier}, \citenamefont
  {Lynn}, \citenamefont {Taniguchi}, \citenamefont {Wakita},\ and\
  \citenamefont {Kadowaki}}]{PhysRevLett.116.217201}%
  \BibitemOpen
  \bibfield  {author} {\bibinfo {author} {\bibfnamefont {H.}~\bibnamefont
  {Takatsu}}, \bibinfo {author} {\bibfnamefont {S.}~\bibnamefont {Onoda}},
  \bibinfo {author} {\bibfnamefont {S.}~\bibnamefont {Kittaka}}, \bibinfo
  {author} {\bibfnamefont {A.}~\bibnamefont {Kasahara}}, \bibinfo {author}
  {\bibfnamefont {Y.}~\bibnamefont {Kono}}, \bibinfo {author} {\bibfnamefont
  {T.}~\bibnamefont {Sakakibara}}, \bibinfo {author} {\bibfnamefont
  {Y.}~\bibnamefont {Kato}}, \bibinfo {author} {\bibfnamefont {B.}~\bibnamefont
  {F\aa{}k}}, \bibinfo {author} {\bibfnamefont {J.}~\bibnamefont {Ollivier}},
  \bibinfo {author} {\bibfnamefont {J.~W.}\ \bibnamefont {Lynn}}, \bibinfo
  {author} {\bibfnamefont {T.}~\bibnamefont {Taniguchi}}, \bibinfo {author}
  {\bibfnamefont {M.}~\bibnamefont {Wakita}},\ and\ \bibinfo {author}
  {\bibfnamefont {H.}~\bibnamefont {Kadowaki}},\ }\bibfield  {title} {\bibinfo
  {title} {{Quadrupole Order in the Frustrated Pyrochlore
  ${\mathrm{Tb}}_{2+x}{\mathrm{Ti}}_{2\ensuremath{-}x}{\mathrm{O}}_{7+y}$}},\
  }\href {https://doi.org/10.1103/PhysRevLett.116.217201} {\bibfield  {journal}
  {\bibinfo  {journal} {Phys. Rev. Lett.}\ }\textbf {\bibinfo {volume} {116}},\
  \bibinfo {pages} {217201} (\bibinfo {year} {2016})}\BibitemShut {NoStop}%
\bibitem [{\citenamefont {Onoda}\ and\ \citenamefont
  {Tanaka}(2011)}]{PhysRevB.83.094411}%
  \BibitemOpen
  \bibfield  {author} {\bibinfo {author} {\bibfnamefont {S.}~\bibnamefont
  {Onoda}}\ and\ \bibinfo {author} {\bibfnamefont {Y.}~\bibnamefont {Tanaka}},\
  }\bibfield  {title} {\bibinfo {title} {{Quantum fluctuations in the effective
  pseudospin-$\frac{1}{2}$ model for magnetic pyrochlore oxides}},\ }\href
  {https://doi.org/10.1103/PhysRevB.83.094411} {\bibfield  {journal} {\bibinfo
  {journal} {Phys. Rev. B}\ }\textbf {\bibinfo {volume} {83}},\ \bibinfo
  {pages} {094411} (\bibinfo {year} {2011})}\BibitemShut {NoStop}%
\bibitem [{\citenamefont {Chung}(2024)}]{chung2024}%
  \BibitemOpen
  \bibfield  {author} {\bibinfo {author} {\bibfnamefont {K.~T.~K.}\
  \bibnamefont {Chung}},\ }\href@noop {} {\bibinfo {title} {{Mapping the Phase
  Diagram of a Frustrated Magnet: Degeneracies, Flat Bands, and Canting Cycles
  on the Pyrochlore Lattice}}} (\bibinfo {year} {2024}),\ \Eprint
  {https://arxiv.org/abs/2411.03429} {arXiv:2411.03429 [cond-mat.str-el]}
  \BibitemShut {NoStop}%
\bibitem [{\citenamefont {Lozano-Gómez}\ \emph
  {et~al.}(2024{\natexlab{b}})\citenamefont {Lozano-Gómez}, \citenamefont
  {Benton}, \citenamefont {Gingras},\ and\ \citenamefont {Yan}}]{lozano2024}%
  \BibitemOpen
  \bibfield  {author} {\bibinfo {author} {\bibfnamefont {D.}~\bibnamefont
  {Lozano-Gómez}}, \bibinfo {author} {\bibfnamefont {O.}~\bibnamefont
  {Benton}}, \bibinfo {author} {\bibfnamefont {M.~J.~P.}\ \bibnamefont
  {Gingras}},\ and\ \bibinfo {author} {\bibfnamefont {H.}~\bibnamefont {Yan}},\
  }\href@noop {} {\bibinfo {title} {{An Atlas of Classical Pyrochlore Spin
  Liquids}}} (\bibinfo {year} {2024}{\natexlab{b}}),\ \Eprint
  {https://arxiv.org/abs/2411.03547} {arXiv:2411.03547 [cond-mat.str-el]}
  \BibitemShut {NoStop}%
\bibitem [{\citenamefont {Benton}(2020)}]{benton2020}%
  \BibitemOpen
  \bibfield  {author} {\bibinfo {author} {\bibfnamefont {O.}~\bibnamefont
  {Benton}},\ }\bibfield  {title} {\bibinfo {title} {{Ground-state phase
  diagram of dipolar-octupolar pyrochlores}},\ }\href
  {https://doi.org/10.1103/PhysRevB.102.104408} {\bibfield  {journal} {\bibinfo
   {journal} {Phys. Rev. B}\ }\textbf {\bibinfo {volume} {102}},\ \bibinfo
  {pages} {104408} (\bibinfo {year} {2020})}\BibitemShut {NoStop}%
\bibitem [{\citenamefont {Pohle}\ \emph {et~al.}(2023)\citenamefont {Pohle},
  \citenamefont {Yamaji},\ and\ \citenamefont
  {Imada}}]{pohle2023groundstates12pyrochlore}%
  \BibitemOpen
  \bibfield  {author} {\bibinfo {author} {\bibfnamefont {R.}~\bibnamefont
  {Pohle}}, \bibinfo {author} {\bibfnamefont {Y.}~\bibnamefont {Yamaji}},\ and\
  \bibinfo {author} {\bibfnamefont {M.}~\bibnamefont {Imada}},\ }\href@noop {}
  {\bibinfo {title} {{Ground state of the {$S$}=1/2 pyrochlore Heisenberg
  antiferromagnet: A quantum spin liquid emergent from dimensional reduction}}}
  (\bibinfo {year} {2023}),\ \Eprint {https://arxiv.org/abs/2311.11561}
  {arXiv:2311.11561 [cond-mat.str-el]} \BibitemShut {NoStop}%
\bibitem [{\citenamefont {Moessner}\ and\ \citenamefont
  {Chalker}(1998{\natexlab{a}})}]{PhysRevB.58.12049}%
  \BibitemOpen
  \bibfield  {author} {\bibinfo {author} {\bibfnamefont {R.}~\bibnamefont
  {Moessner}}\ and\ \bibinfo {author} {\bibfnamefont {J.~T.}\ \bibnamefont
  {Chalker}},\ }\bibfield  {title} {\bibinfo {title} {{Low-temperature
  properties of classical geometrically frustrated antiferromagnets}},\ }\href
  {https://doi.org/10.1103/PhysRevB.58.12049} {\bibfield  {journal} {\bibinfo
  {journal} {Phys. Rev. B}\ }\textbf {\bibinfo {volume} {58}},\ \bibinfo
  {pages} {12049} (\bibinfo {year} {1998}{\natexlab{a}})}\BibitemShut {NoStop}%
\bibitem [{\citenamefont {Moessner}\ and\ \citenamefont
  {Chalker}(1998{\natexlab{b}})}]{PhysRevLett.80.2929}%
  \BibitemOpen
  \bibfield  {author} {\bibinfo {author} {\bibfnamefont {R.}~\bibnamefont
  {Moessner}}\ and\ \bibinfo {author} {\bibfnamefont {J.~T.}\ \bibnamefont
  {Chalker}},\ }\bibfield  {title} {\bibinfo {title} {{Properties of a
  Classical Spin Liquid: The Heisenberg Pyrochlore Antiferromagnet}},\ }\href
  {https://doi.org/10.1103/PhysRevLett.80.2929} {\bibfield  {journal} {\bibinfo
   {journal} {Phys. Rev. Lett.}\ }\textbf {\bibinfo {volume} {80}},\ \bibinfo
  {pages} {2929} (\bibinfo {year} {1998}{\natexlab{b}})}\BibitemShut {NoStop}%
\bibitem [{\citenamefont {Tsvelik}(1992)}]{PhysRevLett.69.2142}%
  \BibitemOpen
  \bibfield  {author} {\bibinfo {author} {\bibfnamefont {A.~M.}\ \bibnamefont
  {Tsvelik}},\ }\bibfield  {title} {\bibinfo {title} {{New fermionic
  description of quantum spin liquid state}},\ }\href
  {https://doi.org/10.1103/PhysRevLett.69.2142} {\bibfield  {journal} {\bibinfo
   {journal} {Phys. Rev. Lett.}\ }\textbf {\bibinfo {volume} {69}},\ \bibinfo
  {pages} {2142} (\bibinfo {year} {1992})}\BibitemShut {NoStop}%
\bibitem [{\citenamefont {Schaden}\ and\ \citenamefont
  {Reuther}(2023)}]{PhysRevResearch.5.023067}%
  \BibitemOpen
  \bibfield  {author} {\bibinfo {author} {\bibfnamefont {Y.}~\bibnamefont
  {Schaden}}\ and\ \bibinfo {author} {\bibfnamefont {J.}~\bibnamefont
  {Reuther}},\ }\bibfield  {title} {\bibinfo {title} {{Bilinear Majorana
  representations for spin operators with spin magnitudes $S > 1/2$}},\ }\href
  {https://doi.org/10.1103/PhysRevResearch.5.023067} {\bibfield  {journal}
  {\bibinfo  {journal} {Phys. Rev. Res.}\ }\textbf {\bibinfo {volume} {5}},\
  \bibinfo {pages} {023067} (\bibinfo {year} {2023})}\BibitemShut {NoStop}%
\bibitem [{\citenamefont {Reuther}\ and\ \citenamefont
  {W\"olfle}(2010)}]{PhysRevB.81.144410}%
  \BibitemOpen
  \bibfield  {author} {\bibinfo {author} {\bibfnamefont {J.}~\bibnamefont
  {Reuther}}\ and\ \bibinfo {author} {\bibfnamefont {P.}~\bibnamefont
  {W\"olfle}},\ }\bibfield  {title} {\bibinfo {title}
  {{{${J}_{1}\text{\ensuremath{-}}{J}_{2}$} frustrated two-dimensional
  Heisenberg model: Random phase approximation and functional renormalization
  group}},\ }\href {https://doi.org/10.1103/PhysRevB.81.144410} {\bibfield
  {journal} {\bibinfo  {journal} {Phys. Rev. B}\ }\textbf {\bibinfo {volume}
  {81}},\ \bibinfo {pages} {144410} (\bibinfo {year} {2010})}\BibitemShut
  {NoStop}%
\bibitem [{\citenamefont {Metzner}\ \emph {et~al.}(2012)\citenamefont
  {Metzner}, \citenamefont {Salmhofer}, \citenamefont {Honerkamp},
  \citenamefont {Meden},\ and\ \citenamefont
  {Sch\"onhammer}}]{RevModPhys.84.299}%
  \BibitemOpen
  \bibfield  {author} {\bibinfo {author} {\bibfnamefont {W.}~\bibnamefont
  {Metzner}}, \bibinfo {author} {\bibfnamefont {M.}~\bibnamefont {Salmhofer}},
  \bibinfo {author} {\bibfnamefont {C.}~\bibnamefont {Honerkamp}}, \bibinfo
  {author} {\bibfnamefont {V.}~\bibnamefont {Meden}},\ and\ \bibinfo {author}
  {\bibfnamefont {K.}~\bibnamefont {Sch\"onhammer}},\ }\bibfield  {title}
  {\bibinfo {title} {{Functional renormalization group approach to correlated
  fermion systems}},\ }\href {https://doi.org/10.1103/RevModPhys.84.299}
  {\bibfield  {journal} {\bibinfo  {journal} {Rev. Mod. Phys.}\ }\textbf
  {\bibinfo {volume} {84}},\ \bibinfo {pages} {299} (\bibinfo {year}
  {2012})}\BibitemShut {NoStop}%
\bibitem [{\citenamefont {Katanin}(2004)}]{katanin2004}%
  \BibitemOpen
  \bibfield  {author} {\bibinfo {author} {\bibfnamefont {A.~A.}\ \bibnamefont
  {Katanin}},\ }\bibfield  {title} {\bibinfo {title} {{Fulfillment of Ward
  identities in the functional renormalization group approach}},\ }\href
  {https://doi.org/10.1103/PhysRevB.70.115109} {\bibfield  {journal} {\bibinfo
  {journal} {Phys. Rev. B}\ }\textbf {\bibinfo {volume} {70}},\ \bibinfo
  {pages} {115109} (\bibinfo {year} {2004})}\BibitemShut {NoStop}%
\bibitem [{\citenamefont {Sandvik}(2010)}]{10.1063/1.3518900}%
  \BibitemOpen
  \bibfield  {author} {\bibinfo {author} {\bibfnamefont {A.~W.}\ \bibnamefont
  {Sandvik}},\ }\bibfield  {title} {\bibinfo {title} {{Computational Studies of
  Quantum Spin Systems}},\ }\href {https://doi.org/10.1063/1.3518900}
  {\bibfield  {journal} {\bibinfo  {journal} {AIP Conference Proceedings}\
  }\textbf {\bibinfo {volume} {1297}},\ \bibinfo {pages} {135} (\bibinfo {year}
  {2010})}\BibitemShut {NoStop}%
\bibitem [{\citenamefont {Kele\ifmmode~\mbox{\c{s}}\else \c{s}\fi{}}\ and\
  \citenamefont {Zhao}(2022)}]{PhysRevB.105.L041115}%
  \BibitemOpen
  \bibfield  {author} {\bibinfo {author} {\bibfnamefont {A.}~\bibnamefont
  {Kele\ifmmode~\mbox{\c{s}}\else \c{s}\fi{}}}\ and\ \bibinfo {author}
  {\bibfnamefont {E.}~\bibnamefont {Zhao}},\ }\bibfield  {title} {\bibinfo
  {title} {{Rise and fall of plaquette order in the Shastry-Sutherland magnet
  revealed by pseudofermion functional renormalization group}},\ }\href
  {https://doi.org/10.1103/PhysRevB.105.L041115} {\bibfield  {journal}
  {\bibinfo  {journal} {Phys. Rev. B}\ }\textbf {\bibinfo {volume} {105}},\
  \bibinfo {pages} {L041115} (\bibinfo {year} {2022})}\BibitemShut {NoStop}%
\bibitem [{\citenamefont {Iqbal}\ \emph
  {et~al.}(2016{\natexlab{a}})\citenamefont {Iqbal}, \citenamefont {Thomale},
  \citenamefont {Parisen~Toldin}, \citenamefont {Rachel},\ and\ \citenamefont
  {Reuther}}]{PhysRevB.94.140408}%
  \BibitemOpen
  \bibfield  {author} {\bibinfo {author} {\bibfnamefont {Y.}~\bibnamefont
  {Iqbal}}, \bibinfo {author} {\bibfnamefont {R.}~\bibnamefont {Thomale}},
  \bibinfo {author} {\bibfnamefont {F.}~\bibnamefont {Parisen~Toldin}},
  \bibinfo {author} {\bibfnamefont {S.}~\bibnamefont {Rachel}},\ and\ \bibinfo
  {author} {\bibfnamefont {J.}~\bibnamefont {Reuther}},\ }\bibfield  {title}
  {\bibinfo {title} {{Functional renormalization group for three-dimensional
  quantum magnetism}},\ }\href {https://doi.org/10.1103/PhysRevB.94.140408}
  {\bibfield  {journal} {\bibinfo  {journal} {Phys. Rev. B}\ }\textbf {\bibinfo
  {volume} {94}},\ \bibinfo {pages} {140408} (\bibinfo {year}
  {2016}{\natexlab{a}})}\BibitemShut {NoStop}%
\bibitem [{\citenamefont {Iqbal}\ \emph
  {et~al.}(2016{\natexlab{b}})\citenamefont {Iqbal}, \citenamefont {Ghosh},
  \citenamefont {Narayanan}, \citenamefont {Kumar}, \citenamefont {Reuther},\
  and\ \citenamefont {Thomale}}]{PhysRevB.94.224403}%
  \BibitemOpen
  \bibfield  {author} {\bibinfo {author} {\bibfnamefont {Y.}~\bibnamefont
  {Iqbal}}, \bibinfo {author} {\bibfnamefont {P.}~\bibnamefont {Ghosh}},
  \bibinfo {author} {\bibfnamefont {R.}~\bibnamefont {Narayanan}}, \bibinfo
  {author} {\bibfnamefont {B.}~\bibnamefont {Kumar}}, \bibinfo {author}
  {\bibfnamefont {J.}~\bibnamefont {Reuther}},\ and\ \bibinfo {author}
  {\bibfnamefont {R.}~\bibnamefont {Thomale}},\ }\bibfield  {title} {\bibinfo
  {title} {{Intertwined nematic orders in a frustrated ferromagnet}},\ }\href
  {https://doi.org/10.1103/PhysRevB.94.224403} {\bibfield  {journal} {\bibinfo
  {journal} {Phys. Rev. B}\ }\textbf {\bibinfo {volume} {94}},\ \bibinfo
  {pages} {224403} (\bibinfo {year} {2016}{\natexlab{b}})}\BibitemShut
  {NoStop}%
\bibitem [{\citenamefont {Alzate-Cardona}\ \emph {et~al.}(2019)\citenamefont
  {Alzate-Cardona}, \citenamefont {Sabogal-Suárez}, \citenamefont {Evans},\
  and\ \citenamefont {Restrepo-Parra}}]{Alzate19a}%
  \BibitemOpen
  \bibfield  {author} {\bibinfo {author} {\bibfnamefont {J.~D.}\ \bibnamefont
  {Alzate-Cardona}}, \bibinfo {author} {\bibfnamefont {D.}~\bibnamefont
  {Sabogal-Suárez}}, \bibinfo {author} {\bibfnamefont {R.~F.~L.}\ \bibnamefont
  {Evans}},\ and\ \bibinfo {author} {\bibfnamefont {E.}~\bibnamefont
  {Restrepo-Parra}},\ }\bibfield  {title} {\bibinfo {title} {{Optimal phase
  space sampling for Monte Carlo simulations of Heisenberg spin systems}},\
  }\href {https://doi.org/10.1088/1361-648X/aaf852} {\bibfield  {journal}
  {\bibinfo  {journal} {Journal of Physics: Condensed Matter}\ }\textbf
  {\bibinfo {volume} {31}},\ \bibinfo {pages} {095802} (\bibinfo {year}
  {2019})}\BibitemShut {NoStop}%
\bibitem [{\citenamefont {Soldatov}\ \emph {et~al.}(2017)\citenamefont
  {Soldatov}, \citenamefont {Nefedev}, \citenamefont {Komura},\ and\
  \citenamefont {Okabe}}]{SOLDATOV2017707}%
  \BibitemOpen
  \bibfield  {author} {\bibinfo {author} {\bibfnamefont {K.}~\bibnamefont
  {Soldatov}}, \bibinfo {author} {\bibfnamefont {K.}~\bibnamefont {Nefedev}},
  \bibinfo {author} {\bibfnamefont {Y.}~\bibnamefont {Komura}},\ and\ \bibinfo
  {author} {\bibfnamefont {Y.}~\bibnamefont {Okabe}},\ }\bibfield  {title}
  {\bibinfo {title} {{Large-scale calculation of ferromagnetic spin systems on
  the pyrochlore lattice}},\ }\href
  {https://doi.org/https://doi.org/10.1016/j.physleta.2016.12.039} {\bibfield
  {journal} {\bibinfo  {journal} {Physics Letters A}\ }\textbf {\bibinfo
  {volume} {381}},\ \bibinfo {pages} {707} (\bibinfo {year}
  {2017})}\BibitemShut {NoStop}%
\bibitem [{\citenamefont {M\"uller}\ \emph {et~al.}(2017)\citenamefont
  {M\"uller}, \citenamefont {Lohmann}, \citenamefont {Richter}, \citenamefont
  {Menchyshyn},\ and\ \citenamefont {Derzhko}}]{PhysRevB.96.174419}%
  \BibitemOpen
  \bibfield  {author} {\bibinfo {author} {\bibfnamefont {P.}~\bibnamefont
  {M\"uller}}, \bibinfo {author} {\bibfnamefont {A.}~\bibnamefont {Lohmann}},
  \bibinfo {author} {\bibfnamefont {J.}~\bibnamefont {Richter}}, \bibinfo
  {author} {\bibfnamefont {O.}~\bibnamefont {Menchyshyn}},\ and\ \bibinfo
  {author} {\bibfnamefont {O.}~\bibnamefont {Derzhko}},\ }\bibfield  {title}
  {\bibinfo {title} {{Thermodynamics of the pyrochlore Heisenberg ferromagnet
  with arbitrary spin $S$}},\ }\href
  {https://doi.org/10.1103/PhysRevB.96.174419} {\bibfield  {journal} {\bibinfo
  {journal} {Phys. Rev. B}\ }\textbf {\bibinfo {volume} {96}},\ \bibinfo
  {pages} {174419} (\bibinfo {year} {2017})}\BibitemShut {NoStop}%
\end{thebibliography}
\end{document}